\documentclass{aastex63}

\usepackage[version=4]{mhchem}
\usepackage{amsmath,amssymb}
\received{}
\revised{}
\accepted{}
\submitjournal{ApJ}

\shorttitle{Incorporating binding energy distributions}
\shortauthors{Furuya}

\newcommand{\pop}[1]{\langle N_{\rm #1} \rangle}
\newcommand{\num}[1]{n_{\rm #1}}

\newcommand{\edes}{E}
\newcommand{\edescrit}[1]{\mu_{{\rm #1}}}
\newcommand{\ehop}{{E_{{\rm hop}}}}

\newcommand{\refull}{{\mathtt{RE\_FULL}}}
\newcommand{\rembs}{{\mathtt{RE\_PDF}}}
\newcommand{\rehmbs}{{\mathtt{RE\_HPDF}}}
\newcommand{\re}{{\mathtt{RE}}}
\newcommand{\nbin}{N_{\rm bin}}

\begin{document}

\title{A framework for incorporating binding energy distribution in gas-ice astrochemical models}

\correspondingauthor{Kenji Furuya}
\email{furuya@astron.s.u-tokyo.ac.jp}

\author[0000-0002-2026-8157]{Kenji Furuya}
\affiliation{National Astronomical Observatory of Japan, Osawa 2-21-1, Mitaka, Tokyo 181-8588, Japan}
\affiliation{Department of Astronomy, Graduate School of Science, University of Tokyo, Tokyo 113-0033, Japan}



\begin{abstract}
One of the most serious limitations of current astrochemical models with the rate equation (RE) approach is that only a single type of binding site is considered in grain surface chemistry, although laboratory and quantum chemical studies have found that surfaces contain various binding sites with different potential energy depths.
When various sites exist, adsorbed species can be trapped in deep potential sites, increasing the resident time on the surface.
On the other hand, adsorbed species can be populated in shallow sites, activating thermal hopping and thus two-body reactions even at low temperatures, where the thermal hopping from deeper sites is not activated.
Such behavior cannot be described by the conventional RE approach.
In this work, I present a framework for incorporating various binding sites (i.e., binding energy distribution) in gas-ice astrochemical models as an extension of the conventional RE approach.
I propose a simple method to estimate the probability density function for the occupation of various sites by adsorbed species, assuming a quasi-steady state.
By using thermal desorption and hopping rates weighted by the probability density functions, the effect of binding energy distribution is incorporated into the RE approach without increasing the number of ordinary differential equations to be solved.
This method is found to be accurate and computationally efficient and enables us to consider binding energy distribution even for a large gas-ice chemical network, which contains hundreds of icy species.
The impact of the binding energy distribution on interstellar ice composition is discussed quantitatively for the first time.

\end{abstract}

\keywords{astrochemistry --- ISM: molecules}




\section{Introduction}
Astrochemical models have been widely used to predict the molecular evolution in astrophysical objects and to interpret molecular line observations.
Grain surface chemistry plays a key role in interstellar chemistry as well as gas-phase chemistry.
The most abundant molecules with heavy elements, such as \ce{H2O} and \ce{CO2}, are
formed by reactions on grain surfaces, forming ice mantles on top of the dust grains \citep[e.g.,][]{hama13}.
The grain surface chemistry affects the gas-phase composition
through thermal and non-thermal desorption \citep[e.g.,][]{jorgensen20}.

One of the most fundamental parameters to describe the surface chemistry in the interstellar medium (ISM) is the binding energy of adsorbed species on the surface.
Laboratory experiments and quantum chemical calculations have shown that astrochemically relevant surfaces, such as water ice, contain various binding sites with different energy depths, i.e., the binding energy has a distribution \citep[e.g.,][]{amiaud06,he14,doronin15,molpeceres20,minissale22,tinacci22,bovolenta22}.
In the rate equation (RE) approach, which is the most widely used method to numerically investigate the molecular evolution in astrophysical objects, the chemical system is described by a set of ordinary differential equations (ODEs) \citep[e.g.,][]{hasegawa92}.
The RE approach commonly assumes that surfaces contain only a single type of binding site, i.e., the surfaces are characterized by a single binding energy for each species \citep[see][for a review]{cuppen17}.
When various binding sites exist, adsorbed species on a surface can be trapped in binding sites with a deep potential, slowing down the surface diffusion timescale \citep[e.g.,][]{hama12} and increasing the resident time on the surface.
On the other hand, species can be populated in shallow potential sites, activating thermal hopping and thus two-body reactions even at low temperatures, where thermal hopping from deeper sites is not activated.
Such behavior cannot be described by the conventional RE approach.
Moreover, in the conventional RE approach, one has to choose a single "representative" value of the binding energy for each species in some way \citep[see e.g.,][]{minissale22}, and the simulation results can depend on the choice of "representative" values \citep{penteado17,iqbal18}.

There are a few previous attempts to consider multiple binding sites or binding energy distribution within the framework of the RE approach \citep{cuppen11,grassi20}.
In \citet{cuppen11} and \citet{grassi20}, chemical species on different types of binding sites are treated as different species, increasing the number of ODEs to be solved.
\citet{cuppen11} considered two types of binding sites with different potential depths for atomic H, and studied how deep binding sites affect the \ce{H2} formation rate.
\citet{grassi20} considered the binding energy distribution by discretizing the distribution into a finite number of bins (51 at the maximum).
They considered a small surface chemical network, which includes thermal desorption and adsorption of atomic H, \ce{H2O}, and CO and the hydrogenation reaction of CO, in addition to C-H-O gas-phase chemical network. 
Note that the treatment of the surface reaction in \citet{grassi20} is not self-consistent; 
the surface reaction was assumed to occur through the Langmuir–Hinshelwood mechanism,
while the thermal hopping of species from one site to another, which modifies the occupation of sites by species, was neglected.
In theory, discretizing the binding energy distribution into bins would be the most straightforward way as done by the previous studies.
In practice, however, it is difficult to handle a large gas-ice chemical network (containing a few hundred surface species), because the number of ODEs to be solved increases significantly \citep{grassi20}.
This problem becomes even more severe if one would like to consider the dust size distribution in the gas-ice astrochemical simulations;
the dust size distribution plays an important role in the formation of interstellar ice in molecular clouds \citep{pauly16} and the chemistry in protoplanetary disks \citep{gavino21,furuya22c}.
Another numerical difficulty is the increase of the stiffness of the ODEs.
In the conventional RE, one of the fastest processes in the grain surface chemistry is the encounter of reactants (e.g., atomic H) on a surface; its timescale is $\tau_{\rm hop}/\Theta$, where $\tau_{\rm hop}$ is the hopping timescale from one site to another and $\Theta$ is the surface coverage of the reactants. 
As $\Theta$ is often much less than unity in the ISM, $\tau_{\rm hop}/\Theta \gg \tau_{\rm hop}$. 
On the other hand, to include binding energy distribution, one has to directly consider the hopping from one site to another with the timescale of $\tau_{\rm hop}$ (see Section \ref{sec:hydrogen}).


It is straightforward to include binding energy distribution in the microscopic kinetic Monte-Carlo (kMC) approach \citep[e.g.,][]{chang05,cuppen07,garrod13,chang14,cazaux17,furuya22a,zhao22}.
However, the microscopic kMC approach for simulating the gas-ice chemistry in the ISM is computationally expensive, because the size of the timestep is limited by the fastest chemical event;
for example, when the activation energy of thermal hopping for atomic H is 250 K \citep{hama12} and the temperature of 10 K, the timescale of the thermal hopping from one site to another is less than 0.1 sec, which is extremely smaller than the timescale of ice formation ($\gtrsim$10$^5$ yr at the gas density of 10$^{4}$ cm$^{-3}$).
Then it is desired to develop a novel method, that enables us to include binding energy distribution in astrochemical models in a computationally cheap way.

In this work, I propose a novel framework for incorporating binding energy distribution in the RE approach, by using thermal desorption and hopping rates weighted by probability density function (PDF) for the occupation of binding sites by chemical species.
The proposed framework is computationally efficient and enables us to consider binding energy distribution even for a large gas-ice chemical network, which contains hundreds of different icy species.
The rest of this work is organized as follows.
In section \ref{sec:hydrogen}, a new framework, labeled $\rembs$, is introduced using a simple hydrogen system as an example.
In section \ref{sec:complicated}, $\rembs$ is adapted to a more complex surface chemical network, where multiple different species and reactions with an activation energy barrier are considered.
In section \ref{sec:garrod}, $\rembs$ is adapted to a large gas-ice chemical network, where hundreds of surface species are included, to investigate the impact of binding energy distribution on the ISM chemistry.
In Section \ref{sec:discuss}, the formation of complex organic molecules, the assumption in $\rembs$, and simpler methods than $\rembs$ are discussed.
Finally, conclusions are presented in Section \ref{sec:conclusion}.

\section{Hydrogen system} \label{sec:hydrogen}
\subsection{Formulation}
Firstly, let us consider a very simple chemical system, which includes adsorption and thermal desorption of atomic H and the \ce{H2} formation on a surface via the recombination of two H atoms.
In the conventional RE approach, this chemical system is described by the following two equations:
\begin{align}
\frac{d\num{H}}{dt} &= - S_{\rm H} \sigma_{\rm gr} v_{\rm th, \, H} \num{H} \num{gr} + k_{{\rm sub}}\pop{H}\num{gr},\label{eq:hgas1} \\
\frac{d\pop{H}\num{gr}}{dt} &= S_{\rm H} \sigma_{\rm gr} v_{\rm th, \, H} \num{H} \num{gr} - 
k_{{\rm sub}}\pop{H}\num{gr} - 2k_{\rm hop} \pop{H}^2 \num{gr}/N_{\rm site}, \label{eq:hice}
\end{align}
where $\num{H}$ is the number density of atomic H in the gas phase, $\pop{H}$ is the average number of atomic H on a single grain, and
$n_{\rm gr}$ is the number density of dust grains in a unit gas volume. 
Then $\pop{H}\num{gr}$ means the number density of atomic H on grains in a unit gas volume.
$S_{\rm H}$ is the sticking probability of atomic H to the surface, $\sigma_{\rm gr}$ is the cross-section of a dust grain, $v_{\rm th, \, H}$ is the thermal velocity of atomic H in the gas phase, and $N_{\rm site}$ is the number of binding sites on a grain surface.
The sublimation ($k_{{\rm sub}}$) and thermal hopping rate coefficients ($k_{{\rm hop}}$) are given by
\begin{align}
&k_{{\rm sub}} = \nu \exp(-\edes_{\rm H}/k_BT), \\
&k_{{\rm hop}} = \nu \exp(-\chi\edes_{\rm H}/k_BT),
\end{align}
where $\nu$ is a pre-exponential factor and assumed to be 10$^{12}$ s$^{-1}$ throughout this work 
unless otherwise stated \citep[but see][]{minissale22,ligterink23}, $k_B$ is the Boltzmann constant, and $T$ is the surface temperature.
$\edes_{\rm H}$ is the binding energy of atomic H on a surface, while $\chi \edes_{\rm H}$ represents the activation energy for surface hopping, where $\chi$ is the hopping-to-binding energy ratio.
Throughout this work, $\chi$ is assumed to be a constant, irrespective of the adsorbed species, although laboratory studies have shown that the value of $\chi$ depends on species \citep{furuya22b}.

When the binding energy distribution of atomic H is considered, the above simple equations become complicated as explained below.
Throughout this work, it is assumed that each binding site allows only one atom/molecule to be adsorbed at most.
I denote the distribution of binding sites for atomic H with the potential energy depth of $\edes_{\rm H}$ as $g_{\rm H}(\edes_{\rm H})$, which satisfies
\begin{align}
\int g_{\rm H}(\edes_{\rm H})d\edes_{\rm H} = 1. \label{eq:g_close}
\end{align}
I denote the fraction of binding sites with $\edes_{\rm H}$, which are occupied by atomic H as $\theta_{\rm H}(\edes_{\rm H})$.
Then the surface coverage by atomic H ($\Theta_{\rm H}$) is defined as
\begin{align}
\Theta_{\rm H} = \int \theta_{\rm H}(\edes_{\rm H}) g_{\rm H}(\edes_{\rm H})d\edes_{\rm H}. \label{eq:Theta_H}
\end{align}
Note that $\Theta_{\rm H} = \pop{H}/N_{\rm site}$.
The equations that describe the adsorption and thermal desorption of atomic H and the \ce{H2} formation are given as following \citep[cf.][]{li10,furuya19}:
\begin{align}
\frac{d\num{H}}{dt} &= -(1-\Theta_{\rm H})S_{\rm H}\sigma_{\rm gr} v_{\rm th, \, H} \num{H} n_{\rm gr} + N_{\rm site} \num{gr} \int k_{\rm sub}(\edes_{\rm H}) \theta_{\rm H}(\edes_{\rm H})g_{\rm H}(\edes_{\rm H}) d\edes_{\rm H} \label{eq:hgas2}, \\
\frac{d\theta_{\rm H}(\edes_{\rm H})}{dt} &= (1-\theta_{\rm H}(\edes_{\rm H})) S_{\rm H} \frac{\sigma_{\rm gr}}{N_{\rm site}}v_{\rm th, \,H} \num{H} - k_{\rm sub}(\edes_{\rm H})\theta_{\rm H}(\edes_{\rm H})  \label{eq:cov_h2} \\ 
&- \int k_{\rm hop}(\edes_{\rm H} \rightarrow \edes_{\rm H}') \theta_{\rm H}(\edes_{\rm H}) [1-\theta_{\rm H}(\edes_{\rm H}')]g_{\rm H}(\edes_{\rm H}')d\edes_{\rm H}' \nonumber \\
&+  [1-\theta_{\rm H}(\edes_{\rm H})]\int k_{\rm hop}(\edes_{\rm H}' \rightarrow \edes_{\rm H}) \theta_{\rm H}(\edes_{\rm H}')g_{\rm H}(\edes_{\rm H}')d\edes_{\rm H}' \nonumber \\
&- \int k_{\rm hop}(\edes_{\rm H} \rightarrow \edes_{\rm H}') \theta_{\rm H}(\edes_{\rm H}) \theta_{\rm H}(\edes_{\rm H}')g_{\rm H}(\edes_{\rm H}')d\edes_{\rm H}' \nonumber \\
&- \theta_{\rm H}(\edes_{\rm H})\int k_{\rm hop}(\edes_{\rm H}' \rightarrow \edes_{\rm H}) \theta_{\rm H}(\edes_{\rm H}')g_{\rm H}(\edes_{\rm H}')d\edes_{\rm H}'. \nonumber
\end{align}
The first term in Eq. \ref{eq:cov_h2} represents adsorption to the surface, while the second term represents thermal desorption.
The factor $[1-\theta_{\rm H}(\edes_{\rm H})]$ is considered, as only one H atom is allowed to be adsorbed per binding site, neglecting the \ce{H2} formation through Eley-Rideal mechanism.
The third term represents thermal hopping from a binding site with $\edes_{\rm H}$ to a site with $\edes_{\rm H}'$, while the fourth term represents the reverse process.
The fifth and sixth terms represent the \ce{H2} formation through the Langmuir-Hinshelwood mechanism.
In the hopping and \ce{H2} formation terms, it is assumed that sites are randomly distributed on a surface, and all sites are maximally connected for simplicity (See Section \ref{sec:caveat}).
The hopping rate coefficient and the activation energy from a site with the binding energy of $\edes_{\rm H}$ to another site with the binding energy of $\edes_{\rm H}'$ are given by
\begin{align}
k_{{\rm hop}}(\edes_{\rm H} \rightarrow \edes_{\rm H}') &= \nu \exp[-\ehop (\edes_{\rm H} \rightarrow \edes_{\rm H}')/k_BT], \\
\ehop (\edes_{\rm H} \rightarrow \edes_{\rm H}') &= \chi \times {\rm min}(\edes_{\rm H},\,\,\edes_{\rm H}') + {\rm max}(0,\,\,\edes_{\rm H} - \edes_{\rm H}'), \label{eq:ehop}
\end{align}
respectively \citep[][]{cazaux17}.
Note that the hopping rate coefficient obeys the microscopic reversibility,
i.e., $k_{\rm hop}(\edes_{\rm H} \rightarrow \edes_{\rm H}')/k_{\rm hop}(\edes_{\rm H}' \rightarrow \edes_{\rm H}) = \exp[-(\edes_{\rm H}-\edes_{\rm H}')/k_BT]$ \citep{cuppen13}.
When $\edes_{\rm H} = \edes_{\rm H}'$, $\ehop (\edes_{\rm H} \rightarrow \edes_{\rm H}') = \chi \edes_{\rm H}$.
The computational cost for solving Eqs. \ref{eq:hgas2} and \ref{eq:cov_h2} is much higher than that for solving Eqs. \ref{eq:hgas1} and \ref{eq:hice};
for example, if the binding energy distribution is discretized to $N_{\rm bin}$ bins, 
the number of ordinary differential equations to be solved is $N_{\rm bin}$ + 1, and the number of reaction rates to be considered is $\sim N_{\rm bin}^2$.
In addition, the system described by Eqs. \ref{eq:hgas2} and \ref{eq:cov_h2} is stiffer than that described by Eqs. \ref{eq:hgas1} and \ref{eq:hice}, because the former system explicitly includes the hopping rate of atomic H from one site to another, while the latter system only includes the sweeping rate of atomic H over many sites to find another H atom (i.e., $k_{\rm hop}\Theta_{\rm H}$).

I propose a framework to obtain approximate solutions for Eqs. \ref{eq:hgas2} and \ref{eq:cov_h2} rather than directly solving them.
By multiplying both sides of Eq. \ref{eq:cov_h2} by $g_{\rm H}(\edes_{\rm H}) N_{\rm site} n_{\rm gr}$ and integrating with respect to $\edes_{\rm H}$, we get
\begin{align}
\frac{d\pop{H}\num{gr}}{dt} &= (1-\Theta_{\rm H})S_{\rm H}\sigma_{\rm gr} v_{\rm th, \,H} \num{H} \num{gr} - 
\pop{H}\num{gr}\int k_{{\rm sub}}(\edes_{\rm H}) P_{\rm H}(\edes_{\rm H})d\edes_{\rm H} \nonumber \\
&- 2 \frac{\pop{H}^2 \num{gr}}{N_{\rm site}}  \int\int k_{\rm hop}(\edes_{\rm H} \rightarrow \edes_{\rm H}') P_{\rm H}(\edes_{\rm H}) P_{\rm H}(\edes_{\rm H}')d\edes_{\rm H} d\edes_{\rm H}', \label{eq:hicep}
\end{align}
where the relation $\pop{H} = \Theta_{\rm H} N_{\rm site}$ was used.
$P_{\rm H}(\edes_{\rm H})$ describes the probability density function (PDF) for the occupation of binding sites by atomic H on the surface and is defined as
\begin{align}
P_{\rm H}(\edes_{\rm H}) = \frac{\theta_{\rm H}(\edes_{\rm H})g_{\rm H}(\edes_{\rm H})}{\Theta_{\rm H}}, \label{eq:def_pi}
\end{align}
satisfying $\int P_{\rm H}(\edes_{\rm H})d\edes_{\rm H}$ = 1 by definition (see Eq. \ref{eq:Theta_H}).
The hopping terms in Eq. \ref{eq:cov_h2} (the fifth and sixth terms) cancel each other out and the corresponding terms do not appear in Eq. \ref{eq:hicep}.
Note that if $P_{\rm H}(\edes_{\rm H})$ is the delta function, Eq. \ref{eq:hicep} is reduced to Eq. \ref{eq:hice}.
Eqs. \ref{eq:hgas2} and \ref{eq:hicep} are not closed, because $\theta_{\rm H}(\edes_{\rm H})$ and thus $P_{\rm H}(\edes_{\rm H})$ are unknown.
However, if $P_{\rm H}(\edes_{\rm H})$ can be estimated accurately in some way with low computational cost, solving Eqs. \ref{eq:hgas2} and \ref{eq:hicep} should be equivalent to solving Eqs. \ref{eq:hgas2} and \ref{eq:cov_h2}, while the computational cost for solving the former would be cheaper than solving the latter.

\subsection{Estimating a PDF for the occupation of binding sites by atomic H} \label{sec:p_est}
\begin{figure*}[ht!]
\plotone{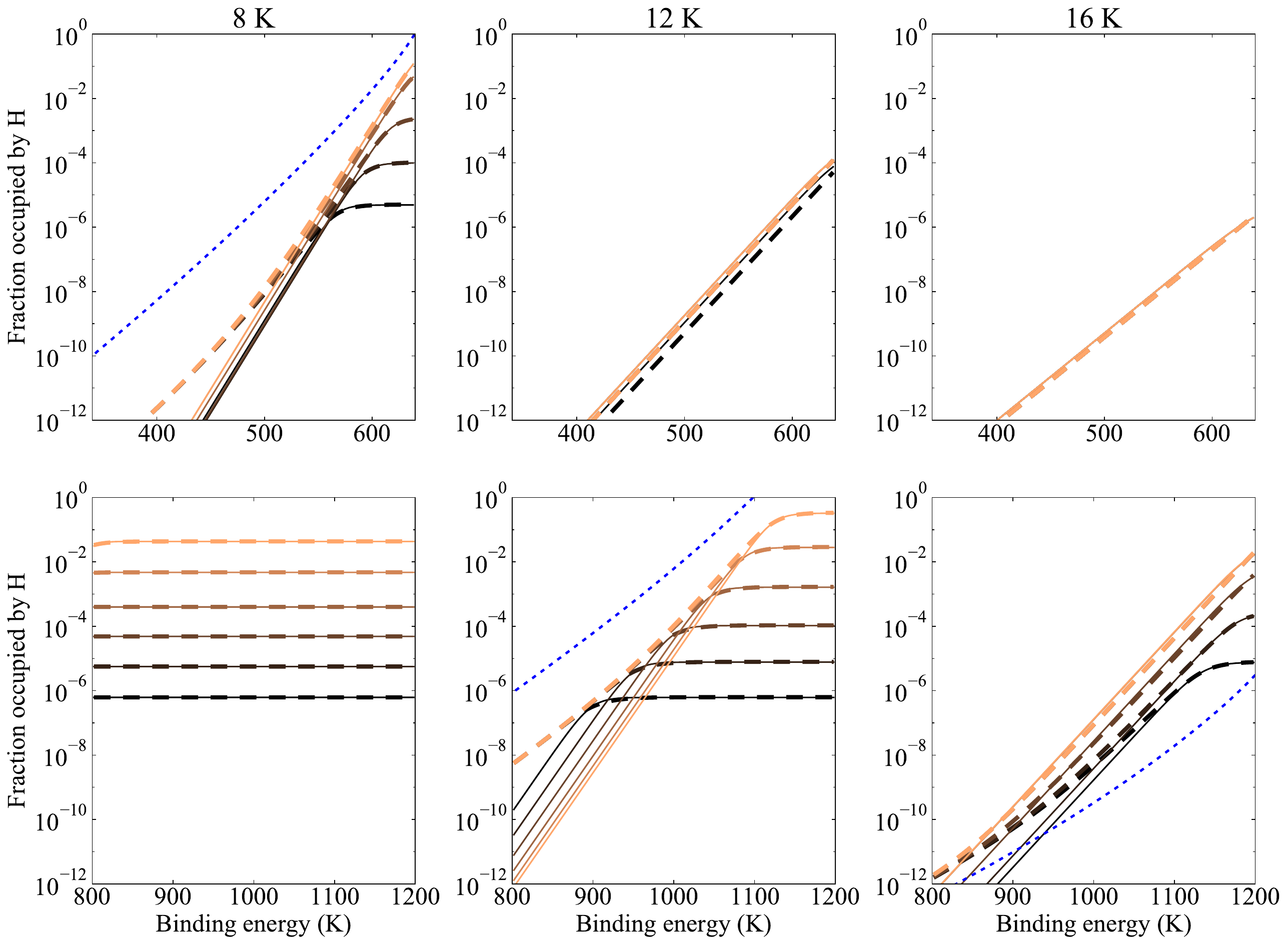}
\caption{Dashed lines show the fraction of binding sites with $\edes_{\rm H}$ occupied by atomic H ($\theta_{\rm H}(\edes_{\rm H})$) at 8 K (left panels), 12 K (middle panels), and 16 K (right panels) obtained by solving Eqs. \ref{eq:hgas2} and \ref{eq:cov_h2}.
Time goes from black lines to orange lines, and the lines are drawn at $10^{-3}$ yr, $10^{-2}$ yr, $10^{-1}$ yr, $1$ yr, $10^{1}$ yr, and $10^{2}$ yr.
Solid lines represent the fit to the numerical results by Eq. \ref{eq:theta_fd} using SciPy's CURVE\_FIT.
Top panels show the cases when the binding energy of atomic H ranges from 240 K to 640 K, while bottom panels show the cases when the binding energy ranges from 800 K to 1200 K.
In both cases, the binding energy distribution follows a Gaussian distribution with a standard deviation of 100 K.
For all the cases, the gas density of atomic H is 2 cm$^{-3}$, and $\chi$ is 0.5.
Blue dotted lines represent the dependence of $\theta_{\rm H}$ on $\edes_{\rm H}$ predicted by Eq. \ref{eq:theta_hop}.
}
\label{fig:fd}
\end{figure*}

A semi-analytical method for estimating $P_{\rm H}(\edes_{\rm H})$ is introduced here.
As $g_{\rm H}(\edes_{\rm H})$ is an input parameter, 
what needs to be done is to estimate $\theta_{\rm H}(\edes_{\rm H})$ in some way.
As a general case, let us consider a surface with $N_{\rm site}$ adsorption sites with the energy depth of $\edes$ that can adsorb one gas atom or molecule on each site.
When the surface is in contact with an ideal gas with a chemical potential of $-\mu_{\rm IG}$, this system can be treated by the grand canonical ensemble in statistical mechanics, and the fraction of occupied sites is given by \citep[Chapter 2 of][]{kubo1988statistical}
\begin{align}
\theta(\edes) = \left(1 + \exp\left(-\frac{\edes-\mu_{\rm IG}}{k_BT}\right)\right)^{-1}. \label{eq:kubo}
\end{align}
The fraction of occupied sites by \ce{H2} in the dense ISM can be well described by Eq. \ref{eq:kubo} (see the Appendix \ref{ap:h2} for more details), likely because the number of \ce{H2} molecules per unit gas volume is much larger than the number of adsorption sites per unit gas volume, and because the adsorption-desorption equilibrium is satisfied for \ce{H2} in the dense ISM conditions \citep{furuya19}.
For species other than \ce{H2}, the adsorption-desorption equilibrium would not be satisfied in general in the ISM, because the adsorption flux is lower than that of \ce{H2} by, 
 at least, several orders of magnitude. 
In addition, the number of atoms/molecules in the gas phase is not always larger than the number of adsorption sites.
Nevertheless, it is empirically found that $\theta_{\rm H}(\edes_{\rm H})$ can be expressed by a similar function to Eq. \ref{eq:kubo}:
\begin{align}
\theta_{\rm H}(\edes_{\rm H}, \edescrit{{\rm H}}) &= C'\left(1 + \exp\left(-\frac{\edes_{\rm H}-\edescrit{{\rm H}}}{k_BT}\right)\right)^{-1}, \label{eq:theta_fd}
\end{align}
where $C'$ is a constant satisfying $C' \leq 1$ and $\edescrit{\rm H}$ is the threshold binding energy ($\edescrit{\rm H} \geq 0$) defined later.
Actually, as shown in Figure \ref{fig:fd}, $\theta_{\rm H}(\edes_{\rm H})$ obtained by solving Eqs. \ref{eq:hgas2} and \ref{eq:cov_h2} is well fitted with Eq. \ref{eq:theta_fd} with $C'$ and $\mu_{\rm H}$ as free parameters.
I confirmed that this is true even in the case when the number of gas-phase atomic H is smaller than the number of adsorption sites.
Then, in the rest of this paper, I assume that $\theta_{\rm H}(\edes_{\rm H})$ at a given time $t$ follows Eq. \ref{eq:theta_fd}.
It is noted that for sites with low $\edes_{\rm H}$ (e.g., $\lesssim$500 K in the top left panel of Fig. \ref{fig:fd}), the dependence of $\theta_{\rm H}$ on $\edes_{\rm H}$ is steeper than predicted by  Eq. \ref{eq:theta_fd}.
For sites with such low $\edes_{\rm H}$, the occupation distribution seems to be determined solely by hopping among sites and to follow (see blue dotted lines in Fig. \ref{fig:fd})
\begin{align}
\theta(\edes_{\rm H}) \propto \tau_{\rm res}(\edes_{\rm H})/\int \tau_{\rm res}(\edes_{\rm H}') g_{\rm H}(\edes_{\rm H}')d\edes_{\rm H}', \label{eq:theta_hop}
\end{align}
where $\tau_{\rm res}(\edes_{\rm H})$ is the resident time on a site with $\edes_{\rm H}$ and is defined as
\begin{align}
\tau_{\rm res}(\edes_{\rm H}) = 1/\int k_{\rm hop}(\edes_{\rm H} \rightarrow \edes_{\rm H}')g_{\rm H}(\edes_{\rm H}')d\edes_{\rm H}'. \label{eq:tau_hop}
\end{align}
The combination of Eq. \ref{eq:theta_fd} and Eq. \ref{eq:theta_hop} may better represent the overall behavior of $\theta_{\rm H}$, but this task is postponed for future work.

Substituting Eq. \ref{eq:theta_fd} to Eq. \ref{eq:def_pi}, the PDF for the occupation of sites by atomic H is rewritten as
\begin{align}
P_{\rm H}(\edes_{\rm H}, \edescrit{{\rm H}}) = C \cdot g_{\rm H}(\edes_{\rm H}) \left(1 + \exp\left(-\frac{\edes_{\rm H}-\edescrit{{\rm H}}}{k_BT}\right)\right)^{-1}, \label{eq:def_p_final}
\end{align}
where $C =(C'/\Theta_{\rm H})$ is a constant and determined from the condition $\int P_{\rm H}(\edes_{\rm H}, \edescrit{{\rm H}})d\edes_{\rm H}$ = 1.
The advantage of this approach is that the PDF for the occupation is characterized by a single parameter, $\mu_{\rm H}$.

The remaining task is to evaluate $\edescrit{H}$.
I estimate $\edescrit{H}$ from the following condition:
\begin{align}
S_{\rm H} \frac{\sigma_{\rm gr}}{N_{\rm site}}v_{\rm th, \,H}(1-\theta_{\rm H}(\edes_{\rm H}, \edescrit{H})) \num{H} = 
\int k_{\rm hop}(\edes_{\rm H} \rightarrow \edes_{\rm H}') \theta_{\rm H}(\edes_{\rm H}, \edescrit{H}) (1-\theta_{\rm H}(\edes_{\rm H}', \edescrit{H})) g_{\rm H}(\edes_{\rm H}')d\edes_{\rm H}'. \label{eq:theta_est}
\end{align}
The left-hand side of the equation represents the adsorption rate of atomic H to sites with $\edes_{\rm H}$, while the right-hand side represents the hopping rate from sites with $\edes_{\rm H}$ to other sites.
The motivation behind Eq. \ref{eq:theta_est} is as follows.
For sites with $\edes_{\rm H} > \edescrit{\rm H}$, $\theta_{\rm H}$ is almost constant regardless of $\edes_{\rm H}$, indicating that adsorption dominates over hopping in those sites (assuming that thermal desorption is slower than hopping).
On the other hand, for sites with $\edes_{\rm H} < \edescrit{\rm H}$, 
$\theta_{\rm H}$ decrease with decreasing $\edes_{\rm H}$;
it can be interpreted that for those sites, thermal hopping dominates over adsorption.
Then it would be reasonable to estimate $\mu_{\rm H}$ by the balance between the adsorption rate and hopping rate.

When $\edescrit{\rm H}$ estimated by Eq. \ref{eq:theta_est} is higher than the maximum value of $\edes_{\rm H}$ (denoted as $\edes_{\rm H,\,max}$), $\edescrit{\rm H}$ is set to $\edes_{\rm H,\,max}$.
On the other hands, if $\int g_{\rm H}(\edes_{\rm H})\theta_{\rm H}(\edes_{\rm H}, \edescrit{\rm H}) d\edes_{\rm H} > \Theta_{\rm H}$ (i.e., there are not enough binding sites with $\edes_{\rm H} \gtrsim \edescrit{\rm H}$ for the occupation of atomic H), $\edescrit{\rm H}$ is determined to satisfy the following relation:
\begin{align}
\int g_{\rm H}(\edes_{\rm H})\theta_{\rm H}(\edes_{\rm H}, \edescrit{\rm H}) d\edes_{\rm H} = \Theta_{\rm H}.
\end{align}

Hereafter, the model which explicitly solves equations for $\theta$ (e.g., Eq. \ref{eq:cov_h2}) is labeled $\refull$, and the approximated method using the PDFs for the occupation of sites is labeled $\rembs$. The conventional RE equation method is labeled $\re$.





\subsection{Numerical test} \label{sec:test1}
To check how well $\rembs$ reproduces the solution obtained with $\refull$, I run a small grid of pseudo-time dependent simulations, where temperature and density are fixed with time in each simulation.
The density of atomic H in the gas phase is set to 2 cm$^{-3}$, which is appropriate in dense molecular gas \citep{goldsmith05}, and gas and dust temperatures are varied from 8 K to 16 K, assuming that the gas and dust temperatures are the same.
The sticking probability of atomic H is set to be unity.
It is assumed that \ce{H2} desorbs into the gas phase upon its formation.
The radius of a dust grain is set to be 0.1 $\mu$m with the dust-to-gas mass ratio of 0.01.
As $g_{\rm H}$, I consider two different cases; in the first case (Case A), $g_{\rm H}$ is assumed to follow a Gaussian distribution with a mean value of 440 K and a standard deviation of 100 K.
The maximum and minimum binding energy are set to be 240 K and 640 K, respectively.
In the second case (Case B), $g_{\rm H}$ is assumed to be the same as that of \ce{D2} and taken from \citet{he14} who determined the binding energy distribution of \ce{D2} on non-porous amorphous water ice by modeling the temperature programmed desorption experiments \citep{amiaud07}.
In both cases, the factor $\chi$ (see Eq. \ref{eq:ehop}) is set to 0.5, and the binding energy distribution is discretized with 100 linearly-spaced bins.
The number of bins is large enough that it does not affect the numerical results.

\begin{figure*}[ht!]
\plotone{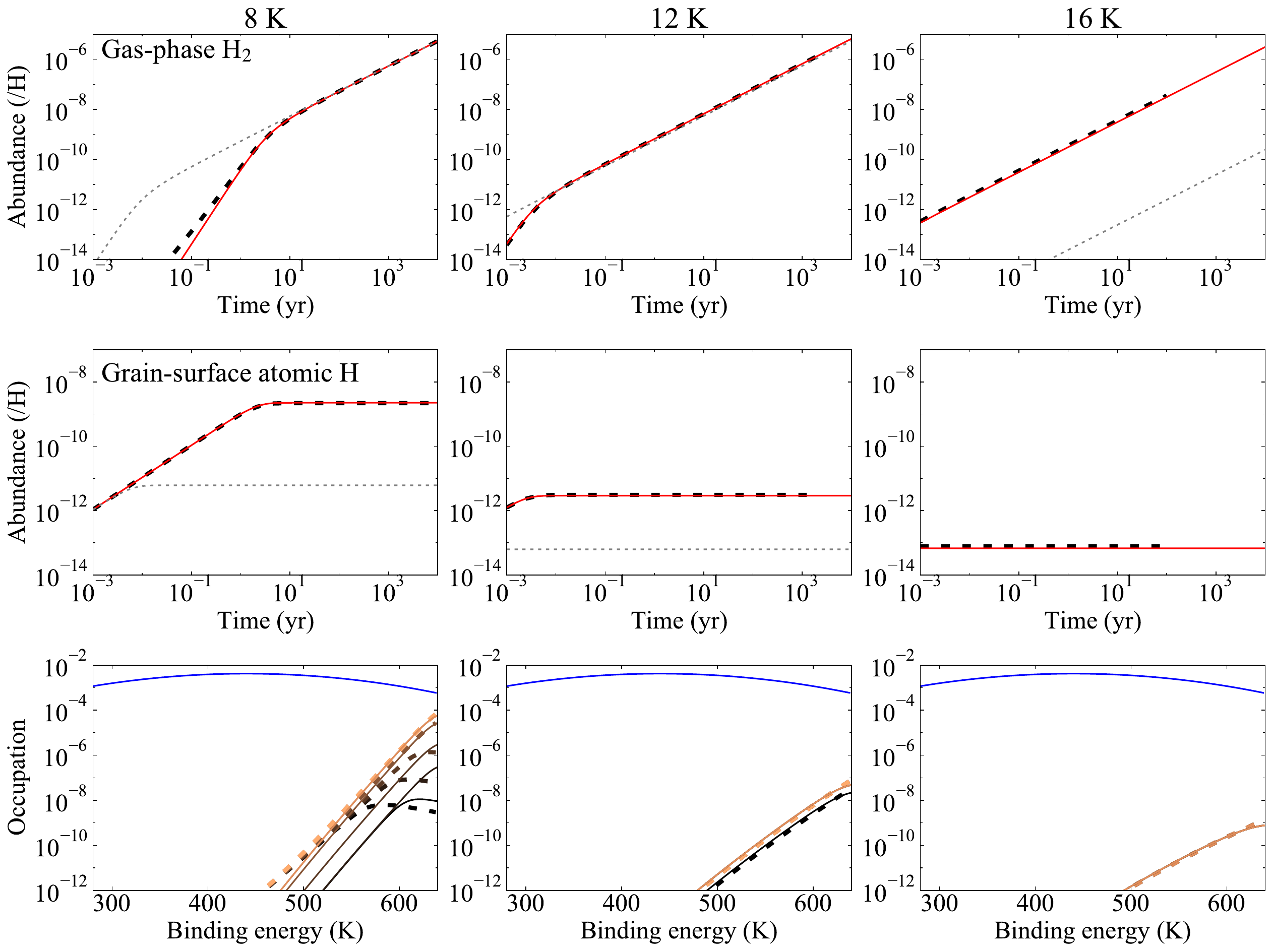}
\caption{Top and middle panels show the temporal evolution of gas-phase \ce{H2} abundance and the abundance of atomic H on surfaces, respectively, predicted by $\rembs$ (red solid lines) and those predicted by $\refull$ (black dashed lines) for Case A.
Gas and dust temperatures are set to be 8 K, 12 K, and 16 K in panels a, b, and c respectively. For comparisons, the dotted gray lines show the prediction by $\re$ using a single binding energy of 440 K.
Bottom panels show the occupation of binding sites by atomic H  ($P_{\rm H}(\edes_{\rm H})\Theta_{\rm H}$) on the surface predicted by $\rembs$ (solid lines) and $\refull$ (dashed lines). 
Time goes from black lines to orange lines, and the lines are drawn at $10^{-3}$ yr, $10^{-2}$ yr, $10^{-1}$ yr, $1$ yr, $10^{1}$ yr, and $10^{2}$ yr.
Blue solid lines show the binding energy distribution $g_{\rm H}$.
}
\label{fig:h2form1}
\end{figure*}

\begin{figure*}[ht!]
\plotone{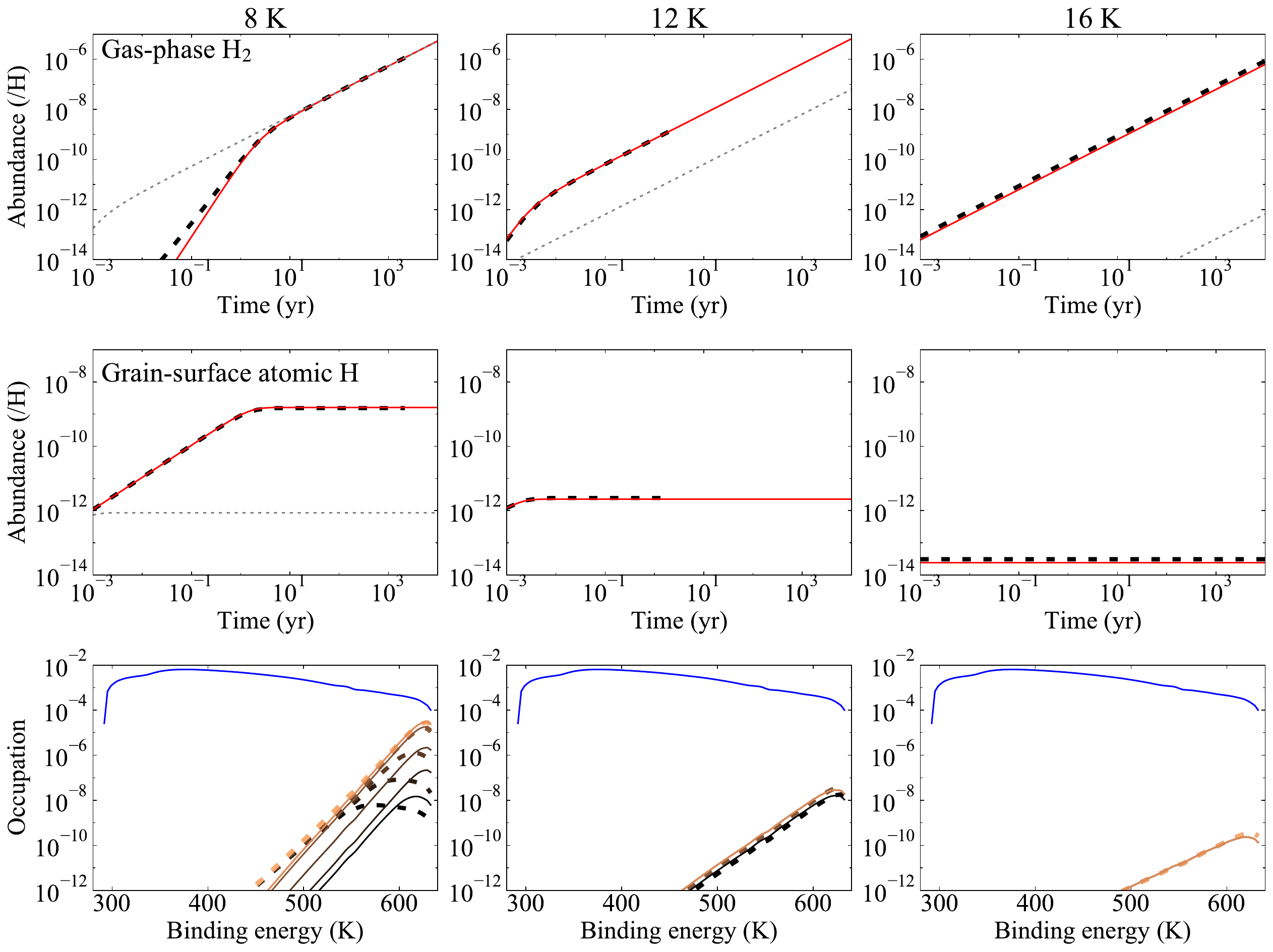}
\caption{Similar to Figure \ref{fig:h2form1}, but for Case B.
The dotted gray lines show the prediction by $\re$ using a single binding energy of 380 K, which corresponds to the peak of the adopted binding energy distribution.
.
}
\label{fig:h2form2}
\end{figure*}

The top and middle panels of Figure \ref{fig:h2form1} compare the abundances of gas-phase \ce{H2} and atomic H on a surface, respectively, predicted by $\refull$ (dashed lines) with those predicted by $\rembs$ (solid lines) in Case A.
It is clear that $\rembs$ reproduces well the abundances of \ce{H2} and atomic H predicted by $\refull$.
For comparisons, the dotted lines show the prediction by $\re$ using a single binding energy of 440 K.
$\rembs$ and $\refull$ predict the higher \ce{H2} formation rate and the higher atomic H abundance on the surface compared to those predicted by $\re$ in particular at 16 K.
The bottom panels show the occupation of binding sites by atomic H ($P_{\rm H}(\edes_{\rm H})\Theta_{\rm H}$).
Atomic H tends to be trapped in deep potential sites ($\gtrsim$600 K), which increases the survival or resident time of atomic H on the surface.
The desorption timescale ($\tau_{\rm des}$) of atomic H from a site with $\edes = 440$ K at 16 K is $\sim$1 second, while the interval at which atomic H is adsorbed on a grain ($\tau_{\rm ads}$) is $>$10$^4$ seconds.
For a site with $\edes \sim 600$ K, $\tau_{\rm des}$ is comparable to $\tau_{\rm ads}$, allowing the \ce{H2} formation.
This explains why $\rembs$ and $\refull$ predict higher \ce{H2} formation rate and the higher atomic H abundance on the surface.
Figure \ref{fig:h2form2} is similar to Figure \ref{fig:h2form1}, but for results of Case B.
Again, $\rembs$ well reproduces the prediction by $\refull$.
 
\section{General formulation} \label{sec:complicated}
\subsection{Formulation and assumption}
Here $\rembs$ is adapted to a more general chemical network, where multiple different species are present on surfaces.
For simplicity, it is assumed that there is no correlation between binding energy for species $i$ and that for species $j$ in a single binding site (i.e., shallow/deep potential sites for species $i$ are not necessarily shallow/deep sites for species $j$), 
and assumed that the PDF for each species is not affected by the presence of other species.
With these assumptions, the PDF for the occupation of sites by each species would be determined independently and is estimated in the same way as presented in Section \ref{sec:p_est}.
The assumption may not be verified when the binding energy of adsorbed species depends on the number of interacting surface species \citep[e.g., see][]{ferrari23} or when different species share the same binding sites.
Nevertheless, these assumptions are employed throughout this paper for simplicity (see also Section \ref{sec:discuss}).

In the framework of $\refull$, the evolutionary equation for $\theta_i(\edes_i)$ through adsorption, thermal desorption, thermal hopping, and two-body reactions are given by
\begin{align}
\frac{d\theta_i(\edes_i)}{dt} & = (1-\theta_i(\edes_i))S_{i} \frac{\sigma_{\rm gr}}{N_{\rm site}}v_{{\rm th},\,i} \num{i} - k_{\rm sub}(\edes_i)\theta_i(\edes_i) \nonumber \\ 
&- \int k_{\rm hop}(\edes_{i} \rightarrow \edes_{i}') \theta_{i}(\edes_{i}) [1-\theta_{i}(\edes_{i}')]g_{i}(\edes_{i}')d\edes_{i}' \nonumber \\
&+  [1-\theta_{i}(\edes_{i})]\int k_{\rm hop}(\edes_{i}' \rightarrow \edes_{i}) \theta_{i}(\edes_{i}')g_{i}(\edes_{i}')d\edes_{i}' \nonumber \\
&+ \Sigma_{j}\Sigma_{k}R^{\rm gain}_{j + k}(\edes_i) - \Sigma_j R_{i + j}^{\rm loss}(\edes_i), \label{eq:theta_gen}
\end{align}
where $R^{\rm gain}_{j+k}$ and $R_{i + j}^{\rm loss}$ are the gain rate and the loss rate, respectively, of species $i$ on sites with $\edes_{i}$ by two body reactions.
When two body reactions are barrierless, they are given by  
\begin{align}
R^{\rm gain}_{k + j}(\edes_i) = \frac{1-\theta_i(\edes_i)}{1-\Theta_i} \left[ \Theta_k \int\int k_{\rm hop}(\edes_j \rightarrow \edes_j') \theta_{j}(\edes_j) (1 - \theta_j(\edes_j'))  g_j(\edes_j)g_j(\edes_j') d\edes_j d\edes_j'  \right. \\
\left. +  \Theta_j \int\int k_{\rm hop}(\edes_k \rightarrow \edes_k') \theta_{k}(\edes_k) (1 - \theta_k(\edes_k'))  g_k(\edes_k)g_k(\edes_k') d\edes_k d\edes_k' \right]\nonumber
\end{align}
and
\begin{align}
R_{i + j}^{\rm loss}(\edes_i) = &- \Theta_j \int k_{\rm hop}(\edes_i \rightarrow \edes_i') \theta_{i}(\edes_i) (1 - \theta_i(\edes_i'))g_i(\edes_i') d\edes_i' \\
&- \theta_i(\edes_i) \int\int k_{\rm hop}(\edes_j \rightarrow \edes_j') \theta_{j}(\edes_j) (1 - \theta_j(\edes_j'))  g_j(\edes_j)g_j(\edes_j') d\edes_j d\edes_j', \nonumber
\end{align}
respectively, again assuming that there is no correlation between binding energy among different species.
The factor ($1-\theta$)/(1-$\Theta$) in the gain rate comes from the assumption that there is no correlation between binding energy for species $i$ and that for species $j$ (or species $j$) in a single binding site.
By multiplying both sides of Eq. \ref{eq:theta_gen} by $g_i(\edes_i) N_{\rm site} n_{\rm gr}$ and integrating with respect to $\edes_i$, we get the general equation of $\rembs$ as
\begin{align}
\frac{d\pop{i}\num{gr}}{dt} &= (1-\Theta_i)S_{i}\sigma_{\rm gr} v_{{\rm th}, \,i} \num{i} \num{gr} - 
K_{i, {\rm sub}}\pop{i}\num{gr} \nonumber \\ &+ \Sigma_{j}\Sigma_{k}\left[\frac{K_{j, {\rm hop}} + K_{k, {\rm hop}}}{N_{\rm site}} \pop{j} \pop{k} n_{\rm gr}\right] - \Sigma_{j}\left[\frac{K_{i, {\rm hop}} + K_{j, {\rm hop}}}{N_{\rm site}} \pop{i} \pop{j} n_{\rm gr}\right].
\label{eq:mbs_general}
\end{align}
The equation can be compared with that of $\re$;
\begin{align}
\frac{d\pop{i}\num{gr}}{dt} &= S_{i}\sigma_{\rm gr} v_{{\rm th}, \,i} \num{i} \num{gr} - 
k_{{\rm sub}}\pop{i}\num{gr} \nonumber \\ &+ 
\Sigma_{j}\Sigma_{k}\left[\frac{k_{{\rm hop}}(\chi\edes_{j}) + k_{{\rm hop}}(\chi\edes_{k})}{N_{\rm site}}\pop{j} \pop{k} n_{\rm gr}\right] - \Sigma_{j}\left[\frac{k_{{\rm hop}}(\chi\edes_{i}) + k_{{\rm hop}}(\chi\edes_{j})}{N_{\rm site}}\pop{i} \pop{j} n_{\rm gr}\right]. \label{eq:re_general}
\end{align}
$K_{i, {\rm sub}}$ and $K_{i, {\rm hop}}$ in Eq. \ref{eq:mbs_general} are defined as
\begin{align}
K_{i, {\rm sub}} = \int k_{{\rm sub}}(\edes_{i}) P_{i}(\edes_{i})d\edes_{i}, \label{eq:kisub}
\end{align}
and
\begin{align}
K_{i, {\rm hop}} = \int\int k_{{\rm hop}}(E_i \rightarrow \edes^{'}_i)
P_i(\edes_i) [g_i(\edes_i')-P_i(\edes_i')\Theta_i] dE_idE_i', \label{eq:kihop}
\end{align}
respectively.
They can be interpreted as averaged sublimation and hopping rate coefficients weighted by the PDF for the occupation.
The factor ($g_i(\edes_i')-P_i(\edes_i')\Theta_i$) comes from the assumption that each binding site allows only one molecule to be adsorbed at most.
If species $i$ can react with itself (e.g., atomic H), that factor should be replaced with $P_i(\edes_i')$ as in Eq. \ref{eq:hicep}.
When $P_{i}$ is the delta function and $\Theta_i \ll 1$, Eq. \ref{eq:mbs_general} is reduced to Eq. \ref{eq:re_general}, 
indicating that $\re$ is a special case of $\rembs$.

In Eq. \ref{eq:mbs_general}, it is assumed that two-body reactions are barrierless.
In $\re$, the rates of barrier-mediated reactions are given by the meeting rate of reactive species multiplied by the factor for describing the reaction-diffusion competition \citep[$f_{{\rm act}}$;][]{chang07,garrod11}:
\begin{align}
f_{ij,\,{\rm act}} = \frac{\nu_{ij} \kappa_{ij,\,{\rm act}}}{\nu_{ij} \kappa_{ij,\,{\rm act}} + k_{{\rm hop}}(\edes_i) + k_{{\rm hop}}(\edes_j)},
\end{align}
where $\nu_{ij}$ is the collisional frequency of the reactants in the binding site and is assumed to be equal to $\nu$ in this work, and $\kappa_{\rm act}$ is the probability of overcoming the activation energy of the reaction per collision.
In $\rembs$, where the hopping rate depends on sites, the rates of barrier-mediated reactions are given by replacing $K_{i, {\rm hop}}$ in Eq. \ref{eq:mbs_general} with $\Gamma_{i+j}$:
\begin{align}
\Gamma_{i+j} = \int\int F_{ij,\,{\rm act}}(E_i') k_{{\rm hop}}(E_i \rightarrow \edes^{'}_i)
P_i(\edes_i) [g_i(\edes_i')-P_i(\edes_i')\Theta_i] dE_idE_i', \label{eq:gamma}
\end{align}
\begin{align}
F_{ij,\,{\rm act}}(E_i') = \frac{\nu_{ij} \kappa_{ij,\,{\rm act}}}
{\nu_{ij} \kappa_{ij,\,{\rm act}} + \int k_{{\rm hop}}(\edes^{'}_i \rightarrow \edes^{''}_i) [g_i(E_i'')-P_i(E_i'')\Theta_i]dE_i'' + 
K_{j, {\rm hop}}
},
\end{align}
where $F_{ij,\,{\rm act}}(E_i')$ describes the reaction-diffusion competition.
Note that $\Gamma_{i+j}$ and $\Gamma_{j+i}$ are not identical.

\subsection{Numerical test} \label{sec:test2}
I consider adsorption and thermal desorption of atomic H, atomic O, and CO, and surface reactions listed in Table \ref{table:reaction}.
For the evaluation of $\kappa_{ij,\,{\rm act}}$, the barriers are assumed to be overcome either thermally or via quantum tunneling assuming rectangular barriers with height $E_{\rm act}$ and width $a$, whichever is faster \citep[e.g.,][]{garrod06}.
For simplifying the simulations with $\refull$, all product species by the surface reactions are assumed to be released to the gas phase immediately (and do not adsorb to the surface again).
The \ce{H2} gas density is set to 10$^4$ cm$^{-3}$.
The gas and dust temperatures are assumed to be the same and set to either 10 K or 15 K.
The binding energy distribution for each species is assumed to follow a Gaussian distribution.
The mean binding energies ($\overline{\edes_i}$) are set to 440 K for atomic H, 1300 K for atomic O, and 1150 K for CO.
A standard deviation ($\sigma_i$) is set to 20 \% of $\overline{\edes_i}$ for each species, and the maximum and minimum binding energies ($\edes_{{\rm max},\,i}$ and $\edes_{{\rm min},\,i}$) are set to $\overline{\edes_i} + 2\sigma_i$ and $\overline{\edes_i} - 2\sigma_i$, respectively.
The binding energy distribution for each species is discretized with 100 linearly-spaced bins.
The parameter $\chi$ for thermal hopping is set to 0.6.
Initially, all atomic H, atomic O, and CO are present in the gas phase with the initial abundances with respect to \ce{H2} of $1.0\times10^{-4}$, $1.8\times10^{-4}$, and $1.0\times10^{-4}$, respectively.
The sticking probability for the three species is assumed to be unity.

\begin{table}
\caption{Surface reactions considered in Section \ref{sec:test2}}
\begin{center}
\begin{tabular}{ccc}
\hline\hline   
Reaction &  $E_{\rm act}$ (K)\tablenotemark{{\rm a}}  & $a$ (\AA)\tablenotemark{{\rm b}}  \\
\hline
H + H $\rightarrow$ \ce{H2}         &  0&  \\  
H + O $\rightarrow$ \ce{OH}         &   0& \\  
O + O $\rightarrow$ \ce{O2}         &  0&  \\  
H + CO $\rightarrow$ \ce{HCO}         & 2500 & 2\\  
O + CO $\rightarrow$ \ce{CO2}         &  1000 & 1\\  
\hline
\end{tabular}
\end{center}
\tablecomments{
$^{a}$Reaction activation energy. 
$^{b}$Barrier width
}
\label{table:reaction}
\end{table}

\begin{figure*}[ht!]
\plotone{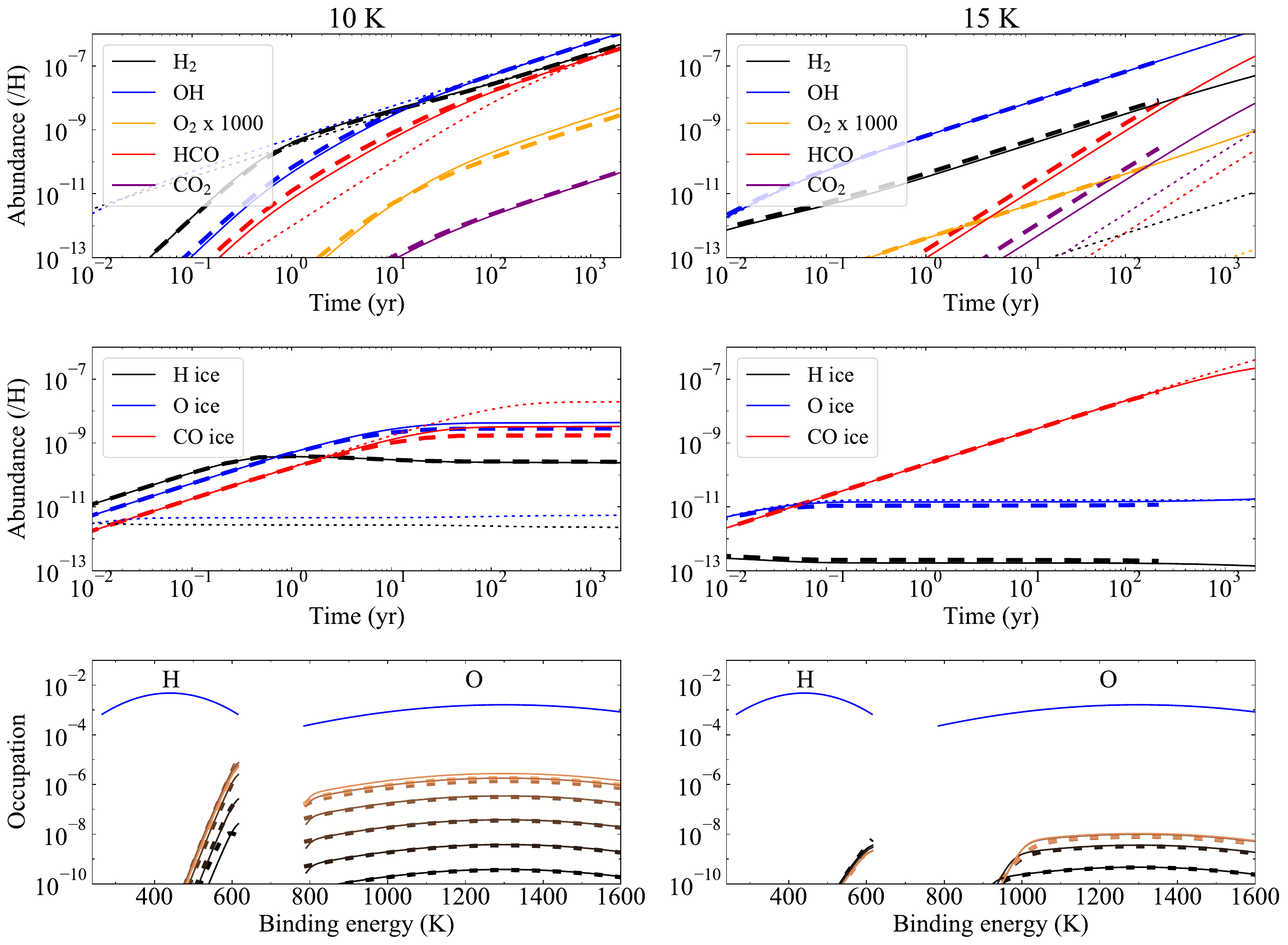}
\caption{Top panels show the gas-phase \ce{H2}, OH, \ce{O2} (enhanced by a factor of 1000 to increase the visibility), HCO and \ce{CO2} abundances predicted by the $\rembs$ model (solid lines) and those predicted by the $\refull$ model (dashed lines). For the \ce{H2} abundance, \ce{H2} produced only by the reaction H + H $\rightarrow$ \ce{H2} is considered in the figures. Middle panels show the abundances of atomic H, atomic O, and CO on grains. Gas and dust temperatures are set to be 10 K (left) or 15 K (right).
For comparisons, the dotted lines show the prediction by $\re$ using a single binding energy (the mean value) for each species.
Bottom panels show the occupation of sites ($P_i(\edes_i)\Theta_i$) by atomic H and atomic O on the surfaces predicted by $\rembs$ (solid lines) and $\refull$ (dotted lines). Time goes from black lines to orange lines, and the lines are drawn at $10^{-3}$ yr, $10^{-2}$ yr, $10^{-1}$ yr, $1$ yr, $10^{1}$ yr, and $10^{2}$ yr. Blue solid lines show the binding energy distribution.
}
\label{fig:oxy}
\end{figure*}

Figure \ref{fig:oxy} shows the temporal evolution of the abundances of the gas-phase molecules (top panels) and the surface species (middle panels) predicted by the $\rembs$ model (solid lines) and by the $\refull$ model (dashed lines).
Both at 10 K and 15 K, the abundances predicted by $\rembs$ and $\refull$ are in good agreement, although there are some deviations (up to a factor of a few).
On the other hand, the abundances predicted by $\rembs$ are significantly different from those predicted by $\re$, indicating that the binding energy distribution plays an important role in the grain surface chemistry.
At 10 K, $\rembs$ predicts several orders of magnitude higher \ce{CO2} and \ce{O2} abundances compared to those predicted by $\re$, 
while the two models predict similar abundances of \ce{H2}, OH, and HCO at $\gtrsim$100 yr.
The formation of \ce{CO2} and \ce{O2} requires the thermal hopping of atomic O and CO.
As shown in the bottom panels, atomic O is populated in the shallowest potential sites with a binding energy of 800 K.
The hopping timescale of atomic O from the shallowest site to another site is $\sim$100 yr (the hopping timescale of CO is smaller than 100 yr as the binding energy of CO is smaller than that of atomic O), which allows the formation of the small amount of \ce{CO2} and \ce{O2} in the $\rembs$ model.
In $\re$, where the binding energy of atomic O is 1300 K, the activation energy for thermal hopping is 780 K; thermal hopping of atomic O is negligible, and thus the formation of \ce{O2} (and \ce{CO2}) is negligible.
The abundances of atomic H and atomic O are considerably higher in $\rembs$ compared to those in $\re$.
In $\rembs$, atomic H tends to be trapped in deep potential sites ($\gtrsim$600 K) and thus the \ce{H2} formation and the hydrogenation reactions are slower, increasing the lifetime of atomic H and O on the surface, compared to that in $\re$.
The higher abundance of atomic O on the surface in $\rembs$ also contributes to the higher production rates of \ce{O2} and \ce{CO2} compared to those in $\re$. 

At 15 K, the prediction by $\re$ is much different from that by $\rembs$ for all species produced by the surface reactions except for OH; the formation rate of OH is mostly determined by the adsorption rate of atomic O both in $\re$ and $\rembs$.
Adsorbed atomic O is populated in sites even with binding energy of $\sim$1000 K and thus its mobility is higher in $\rembs$, leading to more efficient formation of \ce{CO2} and \ce{O2}, compared to those in $\re$.
As atomic H is trapped in sites with $\gtrsim$600 K, the resident time of atomic H on the surface is longer in $\rembs$ than in $\re$.
This allows efficient hydrogenation reactions in $\rembs$ even at 15 K.


\section{Application to a large gas-ice chemical network} \label{sec:garrod}
The main motivation for introducing $\rembs$ is to make it possible to consider binding energy distribution in astrochemical simulations with a large gas-ice chemical network.
Here $\rembs$ is adapted to a slightly modified version of a gas-ice chemical network developed by \citet{garrod13}.
The network includes 709 gas-phase species and 302 ice species connected with $\sim$7000 reactions in total.

\subsection{Simulation setup}
The gas-ice astrochemical code based on $\re$ \citep[Rokko code;][]{furuya15} is updated to be based on $\rembs$.
The chemistry is described by the three-phase model \citep{hasegawa93}, where the gas phase, a surface of ice, and the bulk ice mantle are considered.
As chemical processes, gas-phase reactions, the interaction between gas and (icy) grain surface, and surface reactions, including two-body reactions and photodissociation on the surface, are considered \citep[see][ for details]{furuya15}.
The bulk ice mantle is assumed to be chemically inert.
As non-thermal desorption processes, photodesorption, chemical desorption, and stochastic heating by cosmic-rays are included.
It is assumed that photodesorption yields per incident photon and chemical desorption probabilities per reaction are independent on sites \citep[but see][]{molpeceres23a}.
For the desorption by stochastic heating by cosmic rays, the rate coefficient for each binding site is calculated following the method by \citet{hasegawa93b}, and the rate coefficient for all sites is calculated by taking the average weighted by the PDF for the occupation.
This treatment may overestimate the rate coefficient, because when cosmic rays heat grains, the occupation distribution can change to adjust to increased temperature and species would tend to be populated in deeper sites compared to the occupation distribution before the heating.
Such details are not considered here for simplicity.
In Section \ref{sec:p_est}, the threshold binding energy, $\mu_i$, for estimating $P_i$ is determined from the balance between the hopping rate and the adsorption rate.
Here $\mu_i$ for each species at a given time is determined from the balance between the hopping rate and the production rate, including adsorption, two-body reaction, and photodissociation rates.
As an exception, $\mu_{\ce{H2}}$ is determined from the balance between adsorption and thermal desorption (see Appendix \ref{ap:h2} for details).
In the three-phase model, settling of adsorbing species on a surface is considered, and the rejection terms for adsorption may not be necessary.
It was confirmed that even if the rejection terms ($1-\Theta$ for the adsorption rate in Eq. \ref{eq:theta_est} and $1-\theta$ for the adsorption rate in Eq. \ref{eq:mbs_general}) are omitted, the simulation results presented in this section are unaffected (the differences are less than 1 \%).
This is likely because the surface coverage of reactive species is generally low ($\ll$1).
Hereafter, the rejection terms are omitted.
Exceptionally, the rejection terms for adsorption are needed for \ce{H2}, 
because the surface coverage of \ce{H2} is not negligible \citep{furuya19} and the formation of \ce{H2} multilayer does not occur even at 10 K \citep{gavilan12,kuwahata15}.

The binding energy distribution for each species ($g_i$) is assumed to follow a Gaussian distribution.
Under this assumption, there are four parameters, which characterize the binding energy distribution;
the mean binding energy ($\overline{\edes_i}$), standard deviation ($\sigma_i$), and maximum and minimum binding energies ($\edes_{{\rm max},\,i}$ and $\edes_{{\rm min},\,i}$).
As $\overline{\edes_i}$ for each species, a set of binding energy is taken from \citet{garrod13} and \citet{wakelam17} with some modification.
In particular, the binding energy for atomic H, \ce{H2}, CO, and atomic O are set to 350 K, 350 K, 1150 K, and 1300 K, respectively.
According to quantum chemical calculations of binding energy distribution in the literature, $\sigma_i$ seems to be in the range of $0.1\overline{\edes_i}$-$0.2\overline{\edes_i}$, depending on species \citep{bovolenta22,ferrari23}.
In our models, $\sigma_i$ is set to be either $0.1\overline{\edes_i}$ or $0.2\overline{\edes_i}$.
[$\edes_{{\rm max},\,i}$, $\edes_{{\rm min},\,i}$] are set to be either [$\overline{\edes_i} + 2\sigma_i$, $\overline{\edes_i} - 2\sigma_i$] or [$\overline{\edes_i} + 3\sigma_i$, $\overline{\edes_i} - 3\sigma_i$].
I use a different number of bins for discretizing the binding energy distribution to unify the resolutions of our $\rembs$ models.
For example, in the model with $\sigma_i = 0.2\overline{\edes_i}$ ($\sigma_i = 0.1\overline{\edes_i}$) and [$\edes_{{\rm max},\,i}$, $\edes_{{\rm min},\,i}$] = [$\overline{\edes_i} + 3\sigma$, $\overline{\edes_i} - 3\sigma$] ([$\overline{\edes_i} + 2\sigma$, $\overline{\edes_i} - 2\sigma$]), the number of bins is 100 (33).
The simulation results presented in this section remain essentially the same with more bins (Appendix \ref{sec:comparison}).
The value of $\chi$ for thermal hopping is set to be 0.4 for all species with some exceptions; $\chi$ for atomic H and \ce{H2} are set to be 0.65 \citep{asgeirsson17}, while $\chi$ for CO is set to be 0.3, so that the activation energy for CO hopping is $\sim$350 K \citep{kouchi20}.
$E_{\rm hop}$ for atomic H in the model with $\sigma = 0.2\overline{\edes_{\rm H}}$ is consistent with that constrained by laboratory experiments; 
there are deep potential sites with $E_{\rm hop} \gtrsim 350$ K, while 
the majority of sites are shallow sites with $E_{\rm hop} \lesssim 250$ K on amorphous solid water \citep{hama12}.

I run a small grid of pseudo-time-dependent models with $\rembs$ and $\re$.
The \ce{H2} gas density and the visual extinction ($A_V$) are fixed to be $1\times10^4$ cm$^{-3}$ and 10 mag, respectively.
A temperature range of 10 K to 20 K is considered, assuming the gas and dust temperatures are the same.
Cosmic-ray ionization rate of \ce{H2} is set to be $1.3 \times 10^{-17}$ s$^{-1}$.
The elemental abundances are taken from \citet{aikawa99}.
Initially, all elements are in the form of atoms or atomic ions
except for hydrogen, which is in \ce{H2}.
With Rokko code, the computational time for simulating 1 Myr molecular evolution is typically a few minutes when $\rembs$ is adopted, while it is a few seconds when $\re$ is adopted.
In $\rembs$, the rate coefficients for thermal desorption and two-body surface reactions depend on the abundances through their PDFs (Eqs. \ref{eq:kisub} and \ref{eq:kihop}), and thus it makes the ODE solver \citep[the $\mathtt{LSODES}$ package; ][]{hindmarsh1982odepack} find solutions more difficult compared to the conventional $\re$.
In addition, the computational cost to calculate the rate coefficients for two-body surface reactions is larger by a factor of around $N_{\rm bin}^2$ compared to the conventional $\re$.
These two points are the major reasons for the higher computational cost of $\rembs$ than $\re$.

\subsection{Results} \label{sec:results_repdf}
\begin{figure*}[ht!]
\plotone{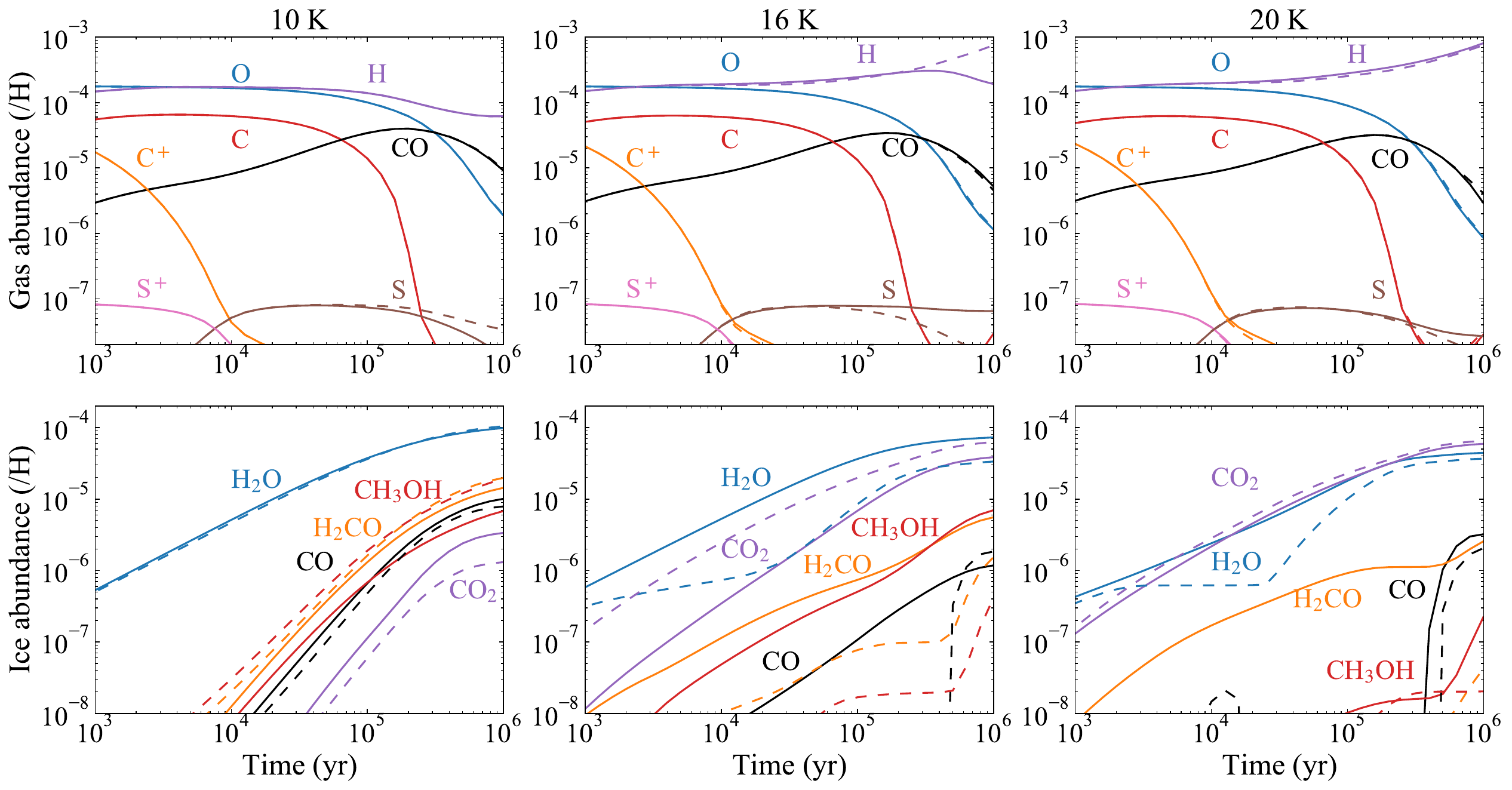}
\caption{Temporal evolution of the abundances of selected gas-phase species (top panels) and selected major icy species (bottom panels) in $\rembs$ (solid lines) and in $\re$ (dashed lines). 
In left, middle, and right panels, gas and dust temperatures are 10 K, 16 K, and 20 K, respectively.
The \ce{H2} gas density and $A_V$ are 10$^4$ cm$^{-3}$ and 10 mag, respectively.
}
\label{fig:full}
\end{figure*}

Figure \ref{fig:full} shows the temporal evolution of selected gas-phase species (top panels) and icy species (bottom panels) with respect to hydrogen nuclei at 10 K (left panels), at 16 K (middle panels), and at 20 K (right panels) in the $\rembs$ model, where $\sigma_i = 0.2\overline{\edes_i}$ and [$\edes_{{\rm max},\,i}$, $\edes_{{\rm min},\,i}$] = [$\overline{\edes_i} + 2\sigma_i$, $\overline{\edes_i} - 2\sigma_i$].
For comparisons, the abundances in the $\re$ model, where the binding energy of each species is set to $\overline{\edes_i}$, are shown by dashed lines.
The abundances of the gas-phase species are almost identical in the $\rembs$ and $\re$ models with some exceptions.
The atomic H abundance in the gas phase is $\sim$10$^{-4}$ both at 10 K and 16 K in the $\rembs$ model, which is determined by the balance between the \ce{H2} ionization rate by cosmic rays and the adsorption rate of atomic H on dust grains \citep{goldsmith05}.
In the $\re$ model at 16 K, adsorbed atomic H is thermally desorbed into the gas phase rather than consumed by surface reactions, leading to higher atomic H abundance than $\sim$10$^{-4}$.
At 20 K, the atomic H abundance is higher than $\sim$10$^{-4}$ both in the $\rembs$ and $\re$ models.
The gas-phase abundance of atomic S, which is the main budget of sulfur in the gas phase at $t\gtrsim10^4$ yr, is slightly different between the $\rembs$ model and the $\re$ model (see the Appendix \ref{sec:detail} for details).
Regarding the icy species, at 10 K, the $\rembs$ model and the $\re$ model predict similar ice abundances and the differences in the predicted abundances are within a factor of a few at $t=10^6$ yr.
At 16 K, the $\rembs$ model predicts that \ce{H2O} ice is the most abundant icy species followed by \ce{CO2} ice.
On the other hand, the $\re$ model predicts that \ce{CO2} ice is more abundant than \ce{H2O} ice.
The lower \ce{CO2} ice abundance in the $\rembs$ model leads to the higher total abundance of CO, \ce{H2CO}, and \ce{CH3OH} ices, the latter two of which are formed through sequential hydrogenation of CO ice, compared to that in the $\re$ model.
The above results indicate that the binding energy distribution plays an important role even in the formation of the major icy species and in the abundances of some gas phase species.
At 20 K, both $\rembs$ and $\re$ models predict that \ce{CO2} ice is more abundant than \ce{H2O} ice.
The abundances in the $\rembs$ and $\re$ models are similar except for \ce{H2CO} ice and \ce{CH3OH} ice;
their abundance are higher in the $\rembs$ model by a factor of $\sim$10 or more than in the $\re$ model.



\begin{figure*}[ht!]
\plotone{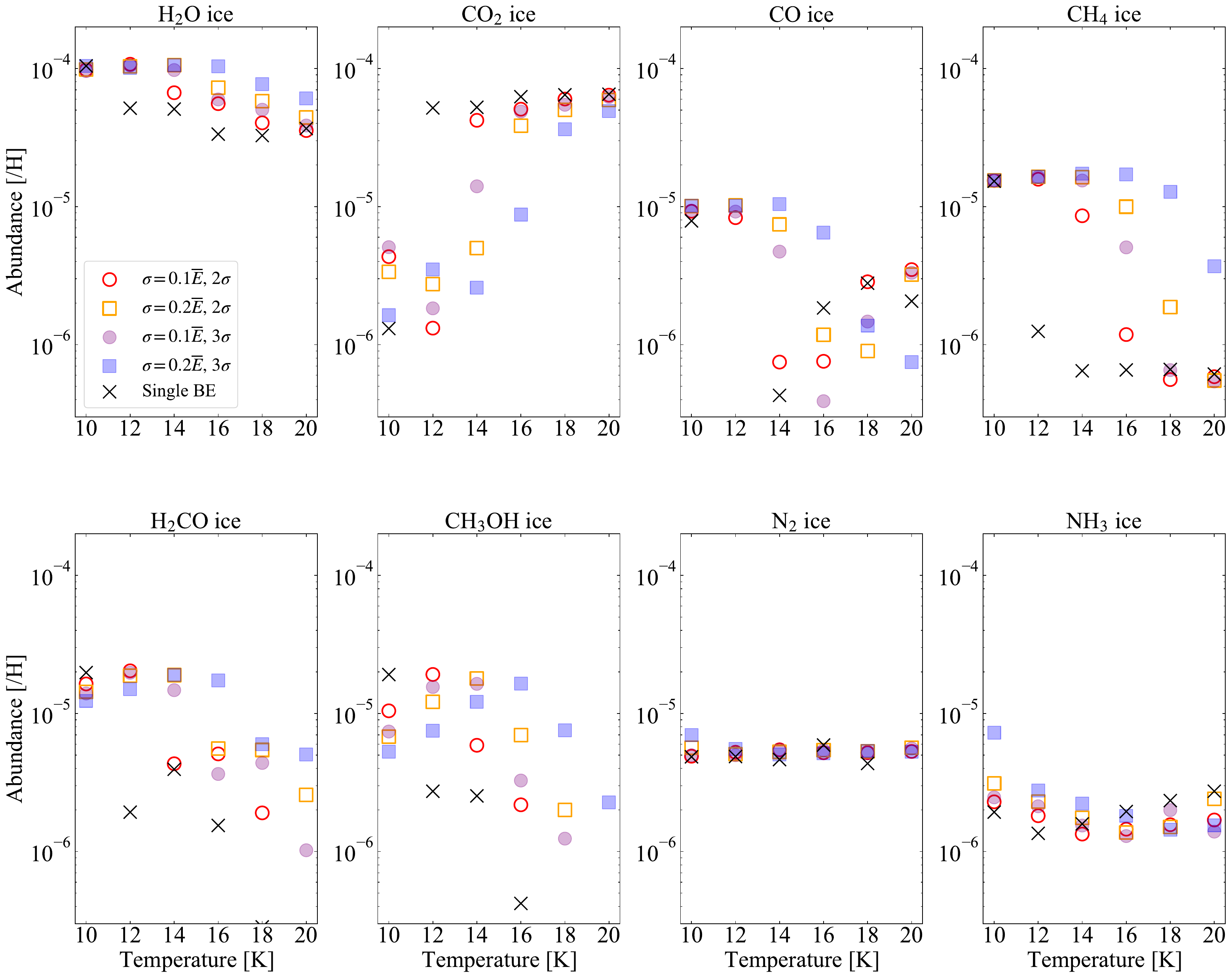}
\caption{Abundances of selected icy species at 10$^6$ yr predicted by $\rembs$ (rectangles and circles) and $\re$ (crosses) at the density of 10$^4$ cm$^{-3}$ and $A_V$ = 10 mag, varying the temperature from 10 K to 20 K. Circles and rectangles represent models where $\sigma_i$ is 0.1$\overline{\edes_i}$ and 0.2$\overline{\edes_i}$, respectively. Open symbols represent the models with [$\edes_{{\rm max},\,i}$, $\edes_{{\rm min},\,i}$] = [$\overline{\edes_i} + 2\sigma_i, \overline{\edes_i} - 2\sigma_i$], while filled symbols represent the models with [$\edes_{{\rm max},\,i}$, $\edes_{{\rm min},\,i}$] = [$\overline{\edes_i} + 3\sigma_i$, $\overline{\edes_i} - 3\sigma_i$].
}
\label{fig:fullgrid}
\end{figure*}

\begin{figure*}[ht!]
\plotone{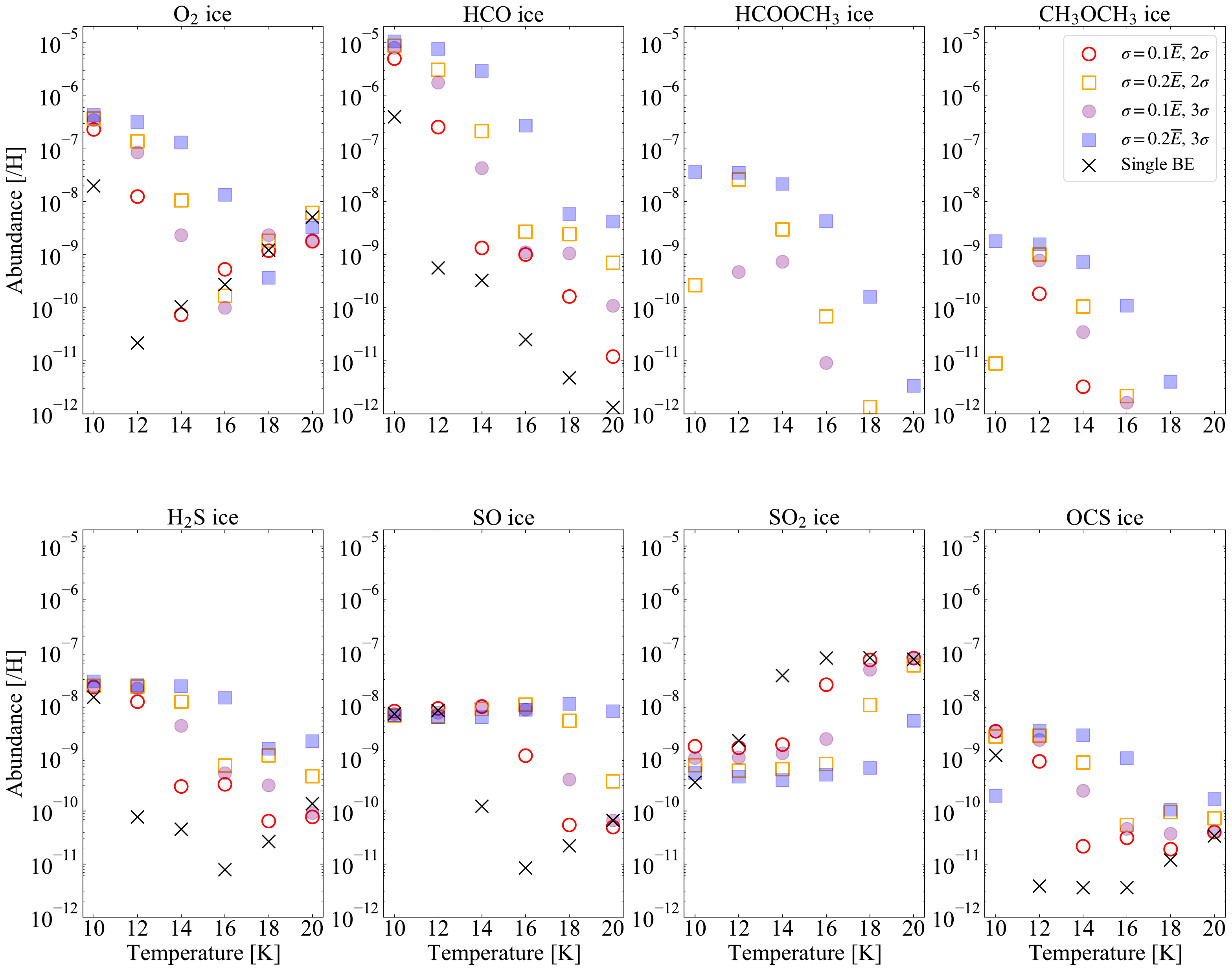}
\caption{Similar to Figure \ref{fig:fullgrid}, but for less abundant species.
}
\label{fig:fullgrid_minor}
\end{figure*}

\begin{figure*}[ht!]
\plotone{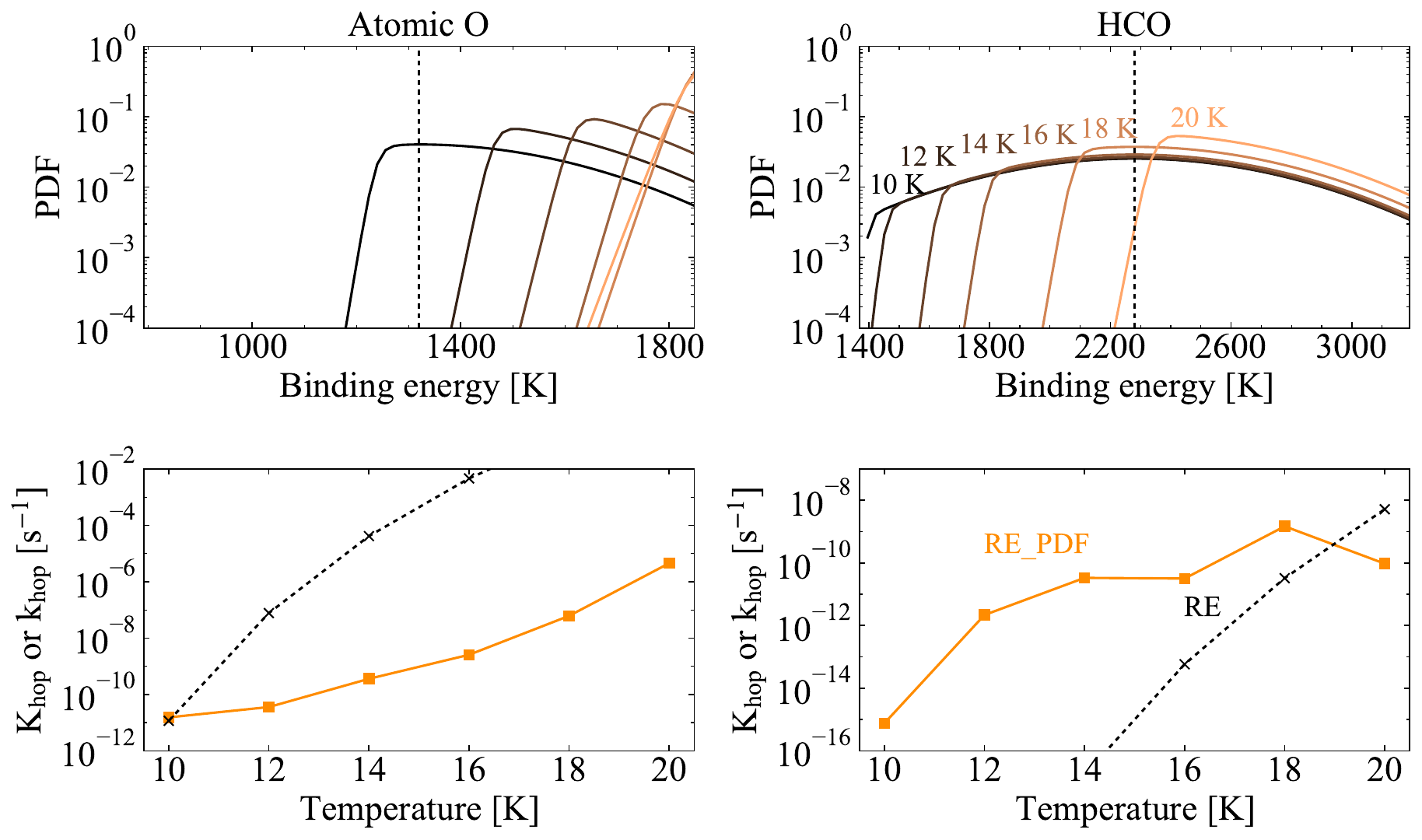}
\caption{
Top panels show the PDF for atomic O (left) and HCO (right) in the $\rembs$ model at 10$^5$ yr, varying in temperature from 10 K to 20 K. The vertical black dotted lines represent the mean binding energy. 
Binding energy distribution is set to [$\edes_{{\rm max},\,i}$, $\edes_{{\rm min},\,i}$] = [$\overline{\edes_i} + 2\sigma_i, \overline{\edes_i} - 2\sigma_i$] and $\sigma_i = 0.2\overline{\edes_i}$.
Bottom panels show $K_{\ce{H},\,\rm hop}$ for atomic O and HCO in the $\rembs$ model (solid orange lines) and $k_{\rm hop}$ for the corresponding species in the $\re$ model (dotted black lines).
}
\label{fig:hop_oxy}
\end{figure*}

Figure \ref{fig:fullgrid} compares the abundances of major icy species at 10$^6$ yr in the $\rembs$ models (rectangles and circles) and in the $\re$ model (black crosses), varying the temperature from 10 K to 20 K.
Typically, the differences in the abundances of the major icy species in the $\rembs$ and $\re$ models are around a factor of 10 or less.
Figure \ref{fig:fullgrid_minor} is similar to Fig. \ref{fig:fullgrid}, but for minor, less abundant icy species.
For those molecules, the differences in the abundances predicted by $\rembs$ and $\re$ reach several orders of magnitude, depending on the assumed binding energy distribution.
As a general trend, the abundances in the $\rembs$ models with higher $\edes_{{\rm max},\,i}$ and lower $\edes_{{\rm min},\,i}$ deviate more significantly from those in the $\re$ model, indicating that the range of the binding energy is important.
Another general trend is that in the models with wider binding energy distributions, the temperature dependence of the ice abundances is weaker.
This trend can be understood as follows.
When $\mu_i > \overline{\edes_i}$, species $i$ tends to be trapped in sites with deeper potential ($\edes_i > \overline{\edes_i}$) than the average one.
On the other hand, when $\mu_i < \overline{\edes_i}$, species $i$ tends to be populated even in shallow ($\edes_i < \overline{\edes_i}$) sites.
In the former case, $K_{i,\,\rm hop}$ is smaller than $k_{\rm hop}$ for the corresponding species, while in the latter case, $K_{i,\,\rm hop}$ can be larger than $k_{\rm hop}$.
As a result, the binding energy distribution tends to make the temperature dependence of the hopping rate coefficient smaller.
As an example, Figure \ref{fig:hop_oxy} shows the PDF and hopping rate coefficient for atomic oxygen and HCO varying in temperature.

\begin{figure*}[ht!]
\plotone{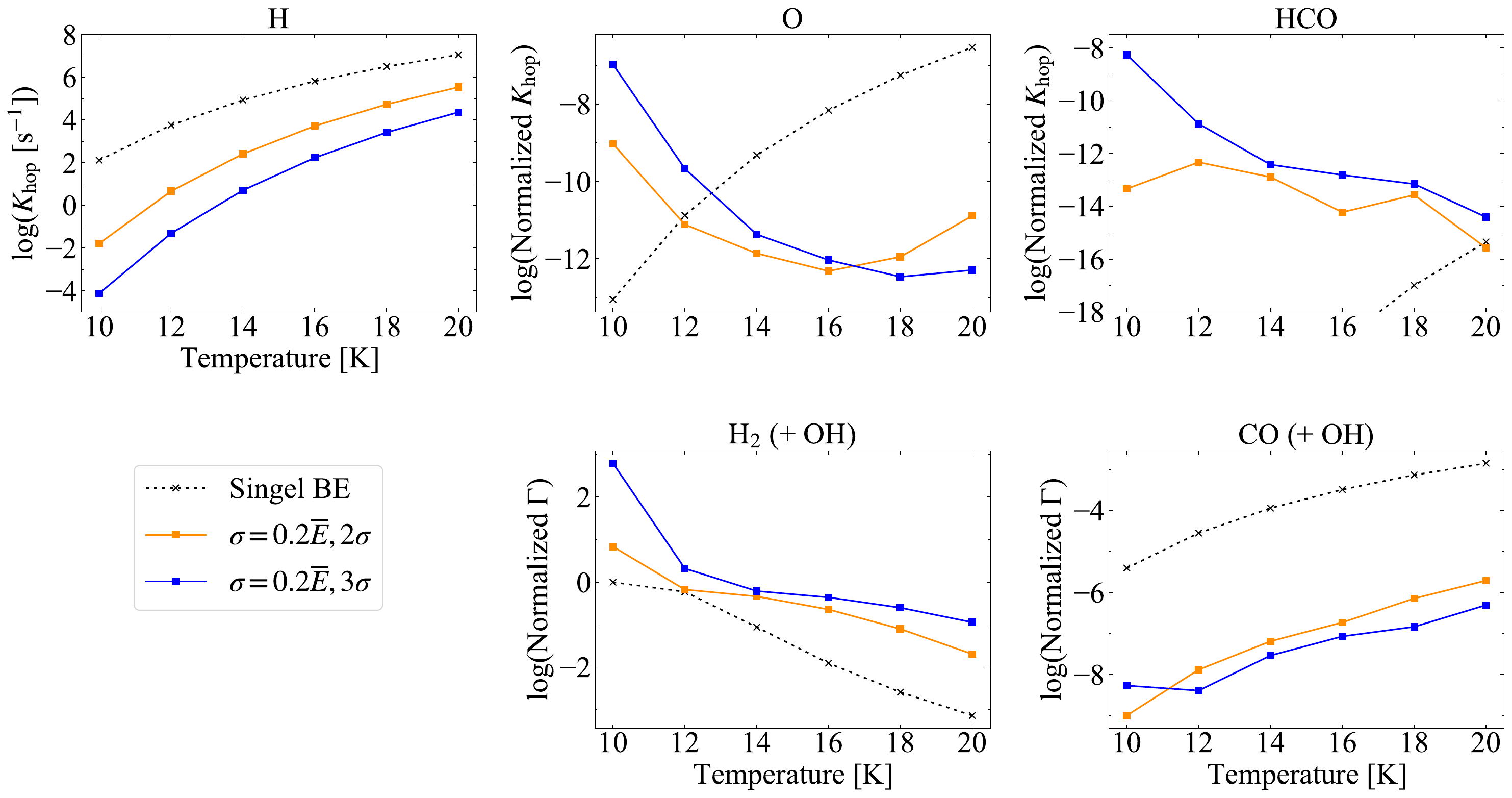}
\caption{Top left panel shows $k_{\rm hop}$ for atomic H in the $\re$ model (black dotted line) and $K_{\ce{H},\,\rm hop}$ in the $\rembs$ model with $\sigma_i = 0.2\overline{\edes_i}$ and [$\edes_{{\rm max},\,i}$, $\edes_{{\rm min},\,i}$] = [$\overline{\edes_i} + 2\sigma_i$, $\overline{\edes_i} - 2\sigma_i$] (orange solid lines) and the $\rembs$ model with $\sigma_i = 0.2\overline{\edes_i}$ and [$\edes_{{\rm max},\,i}$, $\edes_{{\rm min},\,i}$] = [$\overline{\edes_i} + 3\sigma_i$, $\overline{\edes_i} - 3\sigma_i$] (blue solid lines) at $t = 1\times10^5$ yr as functions of the temperature.
Top middle and top right panels show $K_{\ce{O},\,\rm hop}$ and $K_{\ce{HCO},\,\rm hop}$, respectively, normalized by $K_{\ce{H},\,\rm hop}$ in the $\rembs$ models.
Bottom panels show $\Gamma_{\rm \ce{H2}+OH}$ and $\Gamma_{\rm CO+OH}$ normalized by $K_{\ce{H},\,\rm hop}$ in the $\rembs$ models.
Black dotted lines in those panels represent the corresponding normalized hopping rate in the $\re$ model.
}
\label{fig:hop_coeff}
\end{figure*}

The top left panel of Figure \ref{fig:hop_coeff} shows $k_{\rm hop}$ for atomic H in the $\re$ model (black dotted line) and $K_{\ce{H},\,\rm hop}$ in the $\rembs$ models (solid lines) at $t=10^5$ yr as functions of dust temperature.
The choice of $t=10^5$ yr is arbitrary, but it is close to the freeze-out timescale at the gas density of 10$^4$ cm$^{-3}$.
The hopping rate coefficient of atomic hydrogen is orders of magnitude lower in the $\rembs$ models than that in the $\re$ model, because $\mu_{\rm H}$ is $\sim$500 K even at 10 K and is larger than $\overline{\edes_{\rm H}}$ of 350 K.
The smaller hopping rate coefficient of atomic H leads to higher abundances of radicals (e.g., HCO ice in Fig \ref{fig:fullgrid_minor}) in the $\rembs$ models than in the $\re$ model.
The remaining panels in Fig. \ref{fig:hop_coeff} show $K_{\rm hop}$ or $\Gamma$ (see Eq. \ref{eq:gamma}) for atomic O (top middle), HCO (top right), \ce{H2} (bottom middle), and CO (bottom right) normalized by $K_{\rm hop}$ for atomic H along with the corresponding normalized hopping rate coefficient in the $\re$ model.
The purpose of the normalization is to see how the competition between H atom addition reactions and other types of reactions differs between the $\rembs$ models and the $\re$ model.
For \ce{H2} and CO, $\Gamma$ instead of $K_{\rm hop}$ is shown, because the surface reactions involving these molecules often have reaction activation energy (e.g., \ce{H2} + OH $\rightarrow$ \ce{H2O} + H and \ce{CO} + OH $\rightarrow$ \ce{CO2} + H).   
For atomic O, the normalized hopping rate coefficient in the $\rembs$ models is higher than in the $\re$ model at $\le$12 K, while the magnitude relation is opposite at the higher temperatures (see also Fig. \ref{fig:hop_oxy}). 
Such behavior of the atomic O hopping is relevant to e.g., \ce{O2} ice and \ce{SO2} ice.
In the temperature range between 10 K and 20 K, $\mu_{\ce{CO}}$ is higher than $\overline{\edes_{\rm CO}}$.
As a result, CO tends to be trapped in deep sites, leading to a lower normalized hopping rate and thus the lower \ce{CO2} ice abundance in the $\rembs$ models.
In contrast, $\mu_{\ce{HCO}}$ is smaller than $\overline{\edes_{\rm HCO}}$ from 10 K to 18 K, leading to a higher hopping rate of \ce{HCO} (Fig. \ref{fig:hop_oxy}).
The combination of higher \ce{HCO} abundance and higher hopping rate of \ce{HCO} leads to the higher \ce{HCOOCH3} abundance in the $\rembs$ than in the $\re$ model.
Under the assumption that the binding energy distribution for atomic H and \ce{H2} are the same, 
$\mu_{\rm \ce{H2}}$ is lower than $\mu_{\rm \ce{H}}$ (i.e., trapping is less efficient for \ce{H2}), and thus relative importance of reactions involving \ce{H2} is larger in the $\rembs$ models than in the $\re$ model.
Actually, \ce{H2} + OH $\rightarrow$ \ce{H2O} + H is the dominant formation pathway of \ce{H2O} formation and more efficient than OH + H $\rightarrow$ \ce{H2O} and \ce{CO} + OH $\rightarrow$ \ce{CO2} + H at $\le$18 K in the $\rembs$ models.
More details for each species in Fig. \ref{fig:fullgrid_minor} are discussed in Section \ref{sec:coms} and Appendix \ref{sec:detail}.




\section{Discussion} \label{sec:discuss}
\subsection{Formation of Complex Organic Molecules} \label{sec:coms}
\begin{figure*}[ht!]
\plotone{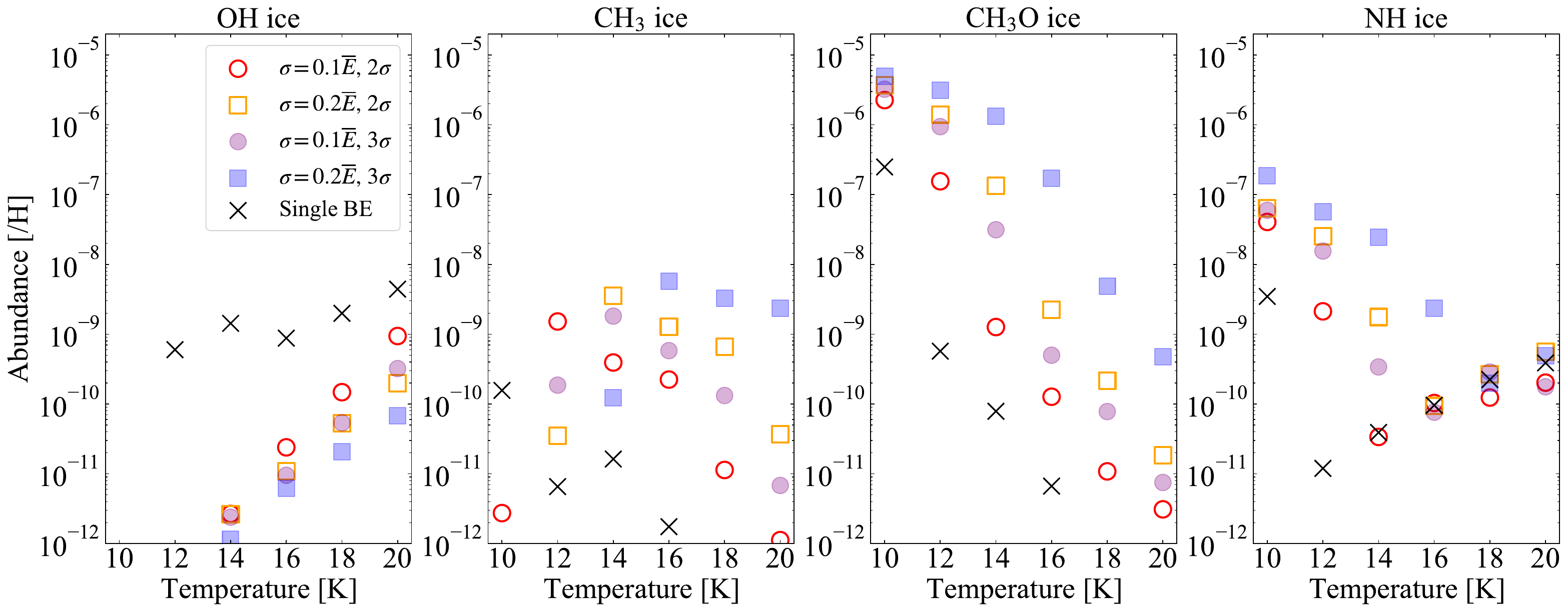}
\caption{Similar to Figure \ref{fig:fullgrid}, but for radicals.
}
\label{fig:fullgrid_radical}
\end{figure*}

As briefly discussed in Section \ref{sec:results_repdf}, the abundances of icy complex organic molecules (COMs; \ce{HCOOCH3} and \ce{CH3OCH3}) in the $\rembs$ model can be several orders of magnitude higher than in the $\re$ model.
The icy COMs are formed by radical-radical reactions on grain surfaces, \ce{HCO} + \ce{CH3O} $\rightarrow$ \ce{HCOOCH3} and \ce{CH3} + \ce{CH3O} $\rightarrow$ \ce{CH3OCH3}, thorough the hopping of \ce{HCO} and \ce{CH3O}.
The mean binding energy for \ce{CH3O} (4400 K) is much higher than that of HCO (2280 K) and \ce{CH3} (1600 K) in the models.
One of the reasons why the icy COMs are abundant in the $\rembs$ is the high abundance of the radicals; the abundance of \ce{HCO} ice and \ce{CH3O} ice in the $\rembs$ model reaches 10$^{-7}$-10$^{-5}$ at the maximum and is higher than that in the $\re$ model by a factor of 10 or more at 10-20 K (Figs. \ref{fig:fullgrid_minor} and \ref{fig:fullgrid_radical}).
\ce{HCO} and \ce{CH3O} are predominantly formed as an intermediate species in the sequence of the hydrogenation of CO ice, which eventually forms \ce{CH3OH} \citep[e.g.,][]{watanabe02b}.
Reflecting slower hydrogenation rates, the abundance of the radicals is higher in the $\rembs$ model compared to the $\re$ model.
Exceptionally, the abundance of radicals that exothermically react with \ce{H2}, such as OH and \ce{CH3}, is not as high as that of HCO and \ce{CH3O} in particular at $\lesssim$16 K.

Furthermore, in contrast to the $\re$ model, the hopping rate coefficient of the radicals is non-negligible even at $\lesssim$16 K in the $\rembs$ models.
The radicals are populated in shallow sites and hop from one site to another (see Fig. \ref{fig:hop_coeff} for \ce{HCO}), and thus \ce{HCOOCH3} and \ce{CH3OCH3} are formed with moderate efficiency even at $\lesssim$16 K via the radical-radical reactions.
As the surface coverage of \ce{CH3O} is a few \% or more at the maximum, a few tens of hopping on average is required for \ce{HCO} to meet \ce{CH3O} and form \ce{HCOOCH3}.
As the timescale of the formation of a monolayer of ice is $\sim$10$^{4}$ yr at the density of $2\times10^4$ cm$^{-3}$, the hopping rate of 10$^{-10}$ s$^{-1}$ is high enough for the efficient formation of \ce{HCOOCH3}.
The hopping rate coefficient of HCO is $\sim$10$^{-12}$-10$^{-11}$ s$^{-1}$ depending on temperature and the binding energy distribution (Fig. \ref{fig:hop_coeff}), and thus a fraction of HCO can react with \ce{CH3O}, but as the abundance of HCO is high, the formation of \ce{HCOOCH3} is non-negligible.
Due to the lower binding energy of \ce{CH3} than \ce{HCO}, the formation of \ce{CH3OCH3} is also non-negligible, although the \ce{CH3} abundance is lower than \ce{HCO}.
These suggest that the enhanced abundance of the icy radicals due to the slower atomic H hopping and the increased hopping rate of the icy radicals play key roles in forming the COMs.
The formation of the icy COMs is favored at lower temperatures, because the radicals are more abundant in $\rembs$.
This trend is the opposite of what is expected from the $\re$ models \citep{garrod06,garrod08}.

The above discussion relies on the assumption that the activation energy of hopping for radicals is relatively low compared to their binding energy ($\chi = 0.4$, Section \ref{sec:garrod}).
The constraint on hopping activation energy of radicals is rather limited at this moment \citep[e.g., ][for the mobility of OH on water ice]{miyazaki22}.
If the hopping activation energy barrier of radicals is higher than assumed in this work, the COMs production at $\lesssim$16 K will be suppressed.
However, the point here is that unlike the previous argument based on $\re$ \citep[e.g.,][]{garrod06,vasyunin13}, the formation of COMs by diffusive radical-radical reactions on surfaces at low temperatures ($\lesssim$16 K) cannot be completely ruled out \citep[see also][]{molpeceres24}.
For more quantitative discussion, information on the mobility of radicals is required and that is beyond the scope of this work. 
Finally, given the high abundance of radicals in ice mantles in $\rembs$ (up to $\sim$10$^{-5}$), the formation efficiency of COMs via the non-diffusive radical-radical reactions in ice mantles may be enhanced by using $\rembs$ compared to the prediction by previous studies with the conventional $\re$ \citep{shingledecker18,jin20,garrod22}.

\subsection{Caveats} \label{sec:caveat}
Some assumptions were employed to construct $\rembs$. 
Firstly, it was assumed that the PDF for the occupation of sites by each species at a given time $t$ is determined solely by the current physical and chemical conditions.
In this work, only pseudo-time-dependent models where the physical conditions are fixed with time are considered.
Under such conditions, the PDF for the occupation would evolve either on the timescale of thermal hopping or adsorption.
If the hopping timescale is shorter than the adsorption timescale, it would be reasonable to assume that the PDF for the occupation is in the steady state.
If the adsorption timescale is shorter than the hopping timescale, the PDF would simply follow the binding energy distribution, and such a situation is considered in the evaluation of the PDF proposed in this work.
Adopting $\rembs$ for the time-dependent physical system would require some modification to the PDF estimation method, in particular, when the dust temperature becomes cooler with time.
When the dust temperature becomes cooler on a shorter timescale than the hopping, the occupation distribution should not change before and after the cooling.
However, in the current method, the occupation distribution immediately adjusts to the new temperature conditions.
Because $\mu$ is lower at lower temperatures, PDF is changed to the direction where more species populate in shallower sites, i.e., in the current method, species trapped in deep sites immediately hop to shallower sites, although the temperature becomes lower and thus the hopping from deep sites to shallower sites becomes less efficient.
In the case where the temperature increases with time, a similar problem can occur. However, as species become more mobile with time, the assumption would be less problematic compared to the cooling case. 

Secondly, it was assumed that the PDF for the occupation of sites by each chemical species is not affected by the presence of other chemical species.
This assumption was employed for the sake of simplicity, and thanks to this simplification, the PDF for each species can be constructed using the information only for each species.
Otherwise, one has to consider the competition for sites between different chemical species, and the PDF for each species should depend on other species.
However, this assumption would not always be verified;
for example, \citet{gavilan12} claimed that the presence of \ce{H2} on grain surfaces enhances the surface diffusion rate of atomic H because deep potential sites can be occupied by \ce{H2}, based on their laboratory experiments.
At this moment, it is unclear how this type of competition affects the ISM chemistry.
To account for the competition between different chemical species, it would be required to use $\refull$ or the microscopic kMC method, although they are much more computationally expensive compared to $\rembs$.

Thirdly, it was assumed that sites with different potential depths are randomly distributed on surfaces, and that all sites are maximally connected.
The connectivity between the different types of sites may be important and play a role in grain surface chemistry \cite[e.g.,][]{cuppen11}. 
Because the rate equation approach, in general, neglects the information on the position of each site, it would not be straightforward to consider non-randomly distributed cases by the rate equation approach with a large chemical network (at least it would make the formulation and coding more difficult). 

In this work, single-peaked binding energy distributions are considered as a first step. However, according to the quantum chemical calculations, the binding energy distribution can be bimodal at least for atomic H and \ce{NH3} on water ice \citep[][]{senevirathne17,tinacci22}. 
For atomic H, I would not expect a significant impact on the shape of the binding energy distribution; 
because the number of H atoms on single grain surface would be much smaller than the number of the binding sites on the surface ($\sim$10$^6$ sites), and because atomic H is mobile on the surface even at $\sim$10 K, only the maximum value of the binding energy would be crucial.
For other species, it is unclear how non-single-peaked distribution affects their chemistry, and whether the $\rembs$ method developed in this work can accurately treat non-single-peaked distributions.
The application of $\rembs$ to non-single-peaked distributions is postponed to future work.

Finally, the modified rate equation method \citep{garrod08,garrod09}, which can take into account the stochastic behavior of grain surface chemistry, is not considered in this work.
It is known that when the number of reactants on a surface is an order of unity or less, the rate equation approach can overestimate the rate of a surface two-body reaction.
Basically, the modified rate equation method replaces the rate of the surface two-body reaction with the adsorption rate of the reactants in the stochastic regime \citep[see][for details and more advanced treatment]{garrod08}.
It is worth noting that in the conventional $\re$, $k_{\rm hop}$ is independent of the adsorption rate coefficient.
On the other hand, in $\rembs$, $K_{\rm hop}$ is determined by referring to the adsorption rate (Eq. \ref{eq:theta_est}).
This does not ensure that $\rembs$ can handle the stochastic behavior of grain surface chemistry without modification.
However, one may expect that the physical-chemical parameter space, where the stochastic behavior of grain surface chemistry appears and surface reaction rates should be modified, is smaller for $\rembs$ compared to the conventional $\re$.
For the stochastic regime, the modified rate method of \citet{garrod08} is probably applicable to the $\rembs$ in a similar way as for the canonical $\re$.
To validate this, the comparison between kMC with binding energy distribution and $\rembs$ would be necessary.

\subsection{Simplified methods} \label{sec:simplified}
One might wonder if it is necessary to consider the binding energy distribution for all species.
Or one might wonder if there is a way to approximate $\rembs$ while using a single value for the binding energy of each species.
Here, two different simpler methods than $\rembs$ are introduced and compared with $\rembs$.
In one model, the binding energy distributions are considered only for atomic H and \ce{H2}, and for the other species, it is assumed that the binding energy has a single fixed value (labeled $\rehmbs$), i.e., the hybrid of $\rembs$ and the conventional $\re$.
In another model, the binding energy of each species has a single binding energy at a given time, but it is a time-variable and set to $\mu$ estimated by the method described in Section \ref{sec:p_est} (labeled $\re\_\mu$).
In $\re\_\mu$, the \ce{H2} abundance on a surface is determined in the same way as in $\rembs$ (Appendix \ref{ap:h2}).
The same chemical network and parameters are used as in Section \ref{sec:results_repdf}.

\begin{figure}[ht!]
\epsscale{0.6}
\plotone{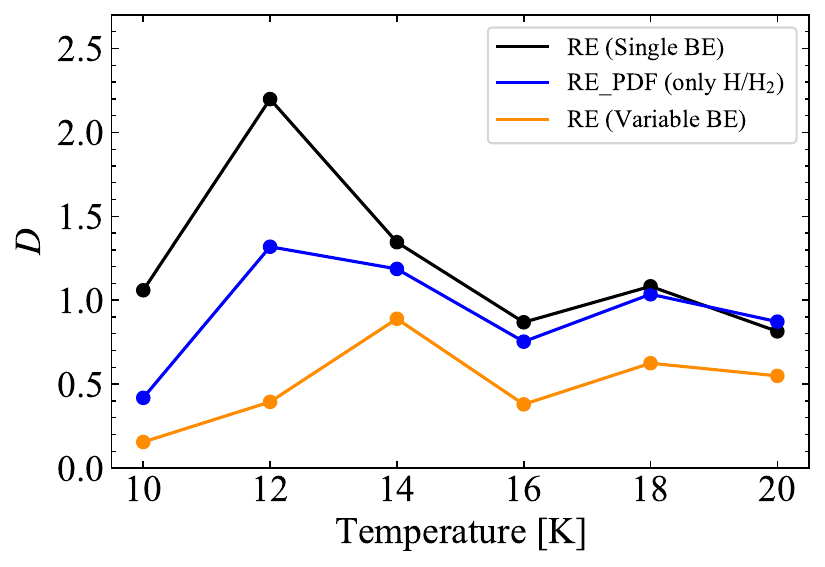}
\caption{
A pseudo-distance (Eq. \ref{eq:dist}) between the $\rembs$ model and the other approximated models discussed in Sect. \ref{sec:simplified}. 
A pseudo-distance between the $\rembs$ model and the conventional $\re$ model is also shown for comparisons.
The gas density is set to 10$^4$ cm$^{-3}$ and $A_V$ is set to 10 mag, varying the temperature from 10 K to 20 K. The abundances at 10$^6$ yr are used for the comparisons.
Binding energy distribution is set to [$\edes_{{\rm max},\,i}$, $\edes_{{\rm min},\,i}$] = [$\overline{\edes_i} + 2\sigma_i, \overline{\edes_i} - 2\sigma_i$] and $\sigma_i = 0.2\overline{\edes_i}$.
}
\label{fig:diff_model}
\end{figure}

It is found that even for species with the abundance of $\gtrsim$10$^{-6}$, the abundances predicted by $\rehmbs$ and $\re\_\mu$ can be different from those predicted by $\rembs$ by a factor of 10 or larger, depending on species and temperature (see Figures \ref{fig:grid_diff} and \ref{fig:grid_diff_minor}).
For obtaining a global view, a pseudo-distance $D$ between the two different models is defined as follows:
\begin{equation}
D = \frac{1}{N_{\rm sp}} \Sigma_i | \log(y^0_i) - \log(\max (y_i, \epsilon)) |, \label{eq:dist}
\end{equation}
where $y^0_i$ and $y_i$ are the abundance of species $i$ predicted by $\rembs$ and that by other method, respectively.
$D=0$ means that the two models give identical results, while $D=1$ means that abundances in the two models differ by a factor of 10 on average. 
$N_{\rm sp}$ is the number of species considered for the comparison, and the icy species with a total (surface + mantle) abundance of larger than 10$^{-12}$ in $\rembs$ are considered.
$N_{\rm sp}$ depended on temperature and ranges from $\sim$140 to $\sim$180.
To avoid significant influence on $D$ from species with a very small abundance, $\epsilon$ is set to 10$^{-15}$.
Figure \ref{fig:diff_model} shows $D$ for $\rehmbs$, $\re\_\mu$, and $\re$ as functions of temperature.
$\rehmbs$ and $\re\_\mu$ give closer results to $\rembs$ compared to $\re$.
In particular, $\re\_\mu$ matches $\rembs$ moderately well and $D$ for $\re\_\mu$ is less than unity at the temperature range from 10 K to 20 K.
$D$ for $\rehmbs$ is considerably smaller than $D$ for $\re$ only at 10 K and 12 K.
Therefore, only considering the binding energy distribution for atomic H and \ce{H2} is not very useful except at 10 K and 12 K.
The moderate success of $\re\_\mu$ indicates that the conventional $\re$ model can be improved by using $\mu_i$ as "representative" binding energy.


\epsscale{1.0}

\section{Conclusion} \label{sec:conclusion}
One of the most serious limitations of current gas-ice astrochemical models with the conventional RE approach is that only a single type of binding site is considered in grain surface chemistry.
Then, one has to choose a single "representative" value of the binding energy for each species, and the simulation results depend on the choice of the "representative" value \citep{penteado17,iqbal18}.
In this work, I introduced a novel framework, $\rembs$, that makes it possible to incorporate the binding energy distribution into gas-ice astrochemical models with the RE approach.
By using the thermal desorption and hopping rates weighted by the PDFs for the occupation of sites by adsorbed species, the effect of binding energy distribution is incorporated into the RE approach without increasing the number of ODEs to be solved.
When the PDF for the occupation is the delta function, $\rembs$ is reduced to the conventional RE approach.
From this point of view, $\rembs$ is a more general framework compared to the conventional RE approach.

It was shown that $\rembs$ can reasonably reproduce the solutions obtained with the more rigorous method ($\refull$) at least for the simple chemical systems (Section \ref{sec:test1} and \ref{sec:test2}).
By adopting $\rembs$ to the large gas-ice chemical network, I showed the importance of the binding energy distribution on the chemistry of dense molecular clouds.
The binding energy distribution plays an important role even in the formation of the major icy species, such as \ce{H2O} ice and \ce{CO2} ice.
As a general trend, in the models with wider binding energy distribution, the abundances of ice molecules show weaker temperature dependence.
The use of $\rembs$ would considerably reduce the uncertainties in gas-ice astrochemical models, which enables us to better understand ISM chemistry through comparisons between the models and observations, such as by the James Webb Space Telescope \citep{yang22,McClure23}.
For that, more laboratory and theoretical studies to quantify binding energy distributions of various species on various types of surfaces are desired.

\acknowledgments
I would like to thank Yuri Aikawa for continuous discussions on astrochemical models, which motivated this work.
I am grateful to Tomoyuki Hanawa for useful comments on the occupation of binding sites, and to Naoki Watanabe, Germ\'{a}n Molpeceres, and Thanja Lamberts for fruitful discussions on the nature of binding energy distributions.
I also thank the anonymous referees for their thorough comments that helped to improve the manuscript considerably.
This work was supported in part by JSPS KAKENHI Grant numbers 20H05847 and 21K13967.

\software{Matplotlib \citep{matplotlib}, SciPy \citep{2020SciPy-NMeth}}




\renewcommand{\thefigure}{A\arabic{figure}}
\setcounter{figure}{0}
\renewcommand{\thetable}{A\arabic{table}}
\setcounter{table}{0}
\begin{appendix}
\section{Special treatment for \ce{H2} on surfaces} \label{ap:h2}
\subsection{\ce{H2} abundance on grain surfaces under adsorption-desorption equilibrium}
In the dense ISM, the adsorption and desorption of \ce{H2} reach the equilibrium in a very short timescale ($\sim$1 yr (10$^4$ cm$^{-3}/n_{\ce{H2}}$)) \citep{furuya19}.
At the adsorption-desorption equilibrium, the occupation of sites by \ce{H2} is determined by the balance 
between the adsorption rate of gaseous \ce{H2} and the thermal desorption rate of adsorbed \ce{H2};
\begin{align}
(1-\theta_{\ce{H2}}(\edes_{\ce{H2}}))S_{\ce{H2}}\num{H_2}\frac{\sigma_{\rm gr}}{N_{\rm site}}v_{\rm th, \,\ce{H2}} = k_{\rm sub}(\edes_{\ce{H2}}) \theta_{\ce{H2}}(\edes_{\ce{H2}}). \label{eq:crit_edes}
\end{align}
By solving Eq. \ref{eq:crit_edes} we obtain \citep[see also][]{amiaud06,furuya19}
\begin{align}
\theta_{\ce{H2}}(\edes_{\ce{H2}}) &= \left(1 + \exp\left(-\frac{\edes_{\ce{H2}}-\mu_{\ce{H2}}}{k_BT}\right)\right)^{-1}, \label{eq:fd_dist} \\
\mu_{\ce{H2}} &= k_B T \ln \left(\frac{4\nu n_{\rm site}}{S_{\ce{H2}}n(\ce{H2})v_{\rm th, \,\ce{H2}}}\right). \label{eq:edes_crit}
\end{align}
For example, at $T = 10 $ K and $n_{\ce{H2}} = 10^4$ cm$^{-3}$, $\mu_{\ce{H2}}$ is $\sim$440 K.
When $S_{\ce{H2}}=1$, $n_{\rm gas} = n_{\ce{H2}}$, and $\nu = (k_B T/h)(2\pi m k_B T/h^2n_{\rm site})$
\footnote{
Based on transition state theory \citep{tait05,minissale22}, neglecting rotational partition function. While $\nu$ is assumed to be constant and 10$^{12}$ s$^{-1}$ in the numerical simulations in this work, $\nu$ based on the transition state theory gives a slightly lower value of $9\times10^{10}$ s$^{-1}$ and $4\times10^{11}$ s$^{-1}$ at 10 K and 20 K, respectively.}, Eq. \ref{eq:fd_dist} is equivalent to Eq. \ref{eq:kubo}, recalling that 
$-\mu_{\rm IG}$ is the chemical potential of ideal gas (Section \ref{sec:p_est}) and is given by 
\begin{equation}
    -\mu_{\rm IG} = k_B T \ln \left[n_{\rm gas}\left(\frac{h^2}{2\pi m k_B T} \right)^\frac{3}{2}\right], \label{eq:myu_ig}
\end{equation}
where $n_{\rm gas}$ and $m$ are the number density of the gas and the mass of each gas particle  \citep{kubo1988statistical}.

Thermal hopping does not affect $\theta_{\ce{H2}}(\edes_{\ce{H2}})$ under adsorption-desorption equilibrium as long as the microscopic reversibility is satisfied \citep{furuya19}.
From $\theta_{\ce{H2}}$, the surface coverage by \ce{H2} is calculated as
\begin{align}
\Theta_{\ce{H2}} = \int \theta_{\ce{H2}}(\edes_{\ce{H2}})g(\edes_{\ce{H2}}) d\edes_{\ce{H2}}. \label{eq:fd_theta}
\end{align}
This equilibrium \ce{H2} coverage would not be changed by the surface two-body reactions that include \ce{H2} as reactants or products, 
given the much higher adsorption rate of \ce{H2} than that of species with heavy elements.
The potential impact of non-thermal desorption on the \ce{H2} coverage is discussed in Section \ref{sec:non-therm}.

In the three-phase astrochemical model based on $\rembs$ (Section \ref{sec:garrod}), 
it is assumed that the \ce{H2} coverage is always given by Eqs. \ref{eq:fd_dist}-\ref{eq:fd_theta}.
In order to fix the \ce{H2} coverage for given physical conditions conserving the number of hydrogen, I modify the ODEs for \ce{H2} in the gas phase and on the surface as follows:
\begin{align}
\left[\frac{dn_{\ce{H2}}}{dt}\right]_{\rm mod} = \left(\frac{dn_{\ce{H2}}}{dt} + \frac{d\pop{\ce{H2}}\num{gr}}{dt}\right) \times \frac{n_{\ce{H2}}}{n_{\ce{H2}} + \pop{\ce{H2}}\num{gr}}, \\
\left[\frac{d\pop{\ce{H2}}\num{gr}}{dt}\right]_{\rm mod} = \left(\frac{dn_{\ce{H2}}}{dt} + \frac{d\pop{\ce{H2}}\num{gr}}{dt}\right) \times \frac{\pop{\ce{H2}}\num{gr}}{n_{\ce{H2}} + \pop{\ce{H2}}\num{gr}},
\end{align}
where $\pop{\ce{H2}}$ in the right-hand side of the equations is given by $\Theta_{\ce{H2}}N_{\rm site}$.

In the three-phase or multi-phase model, molecules on the surface layers become part of the ice mantle with the growth of ice, 
while a part of the ice mantle becomes the surface layers with the loss of ice \citep{hasegawa93,furuya17}.
For evaluating the surface-mantle transition rate, 
one has to calculate the net rate of change in the total surface material (i.e., the difference between the gain rate and the loss rate of the total surface material) at every time step.
In this work, as the adsorption-desorption equilibrium for \ce{H2} is assumed (i.e., the adsorption rate and the desorption rate are always balanced for \ce{H2}), I neglect \ce{H2} in the calculation of the transition terms.

The assumption on the equilibrium \ce{H2} abundance can reduce the numerical difficulty in three-phase and multi-phase models.
In these models, it often happens that the net growth rate is dominated by \ce{H2} adsorption, while the net loss rate is dominated by \ce{H2} thermal desorption.
When the \ce{H2} adsorption and desorption are almost balanced, the net rate of change in the total surface material can be much smaller than the absolute values of \ce{H2} adsorption/desorption rates.
Then the numerical error in the net rate of change can be much larger than the error in the \ce{H2} adsorption/desorption rates (i.e., error in the \ce{H2} gaseous/surface abundances),
which makes it difficult to solve the ODE system \citep[see Sect. 9 of ][]{cuppen17}.
\citet{ruaud16} proposed a different way to calculate the surface-mantle transition rate;
they calculated the transition rate from the surface to ice mantles based on the gain rate of the total surface material, and the transition rate from the ice mantles to the surface based on the loss rate (see their Sect. 2.1).
Their proposed way can reduce the numerical error explained above (because the net rate of change in the total surface material is not directly used in their way) but can cause artificial exchange between surface and bulk ice mantles, making compositions of the surface and bulk ice mantles similar.
The timescale of the artificial exchange would be given by the total amount of ice species divided by the \ce{H2} adsorption rate; assuming the total ice abundance of 10$^{-4}$ and the \ce{H2} gas density of 10$^{4}$ cm$^{-3}$, the timescale is an order of 10 yr, which is much shorter than the timescale of ice formation.
As a result, the method proposed by \citet{ruaud16} with the \ce{H2} adsorption may behave closer to the two-phase model \citep{hasegawa92} rather than the original three-phase model \citep{hasegawa93}.

\subsection{Non-thermal desorption of \ce{H2}} \label{sec:non-therm}
Here the potential impact of non-thermal desorption on the \ce{H2} coverage is discussed.
To the best of my knowledge, the photodesorption yield of \ce{H2} on astronomically relevant surfaces is not quantified in laboratory studies, while there are many measurements for other species, such as water, CO, and \ce{N2}.
These laboratory studies have found that the photodesoption yield per incident FUV photon depends on several parameters, such as adsorbate species, the type of surface, energy of incident photon etc, while the yield is commonly much lower than unity ($\sim$10$^{-4}$-10$^{-2}$) \citep[e.g.,][]{oberg09,hama10,fayolle11,fayolle13,Cruz-Diaz16,paardekooper16,bulak20}.
The FUV flux of interstellar radiation field and that of cosmic-ray (CR) induced photons are $2\times10^{8}$ cm$^{-2}$ s$^{-1}$ and 10$^{4}$ cm$^{-2}$ s$^{-1}$ \citep{shen04}, respectively, 
the latter of which dominates over the former in regions with the visual extinction larger than $\sim$5 mag.
The \ce{H2} flux is $\sim2\times10^{8}(\num{H_2}/10^{4}\,\,{\rm cm}^{-3}$) cm$^{-2}$ s$^{-1}$, which is comparable to the FUV flux of non-attenuated interstellar radiation field 
and is much higher than that of CR induced photons.
Then, photodesorption of \ce{H2} would not affect the surface coverage by \ce{H2} significantly.

Regarding stochastic heating by CR particles, one may consider two types of CR particles: protons and irons \citep[e.g.,][]{leger85,herbst06}.
The former is more abundant than the latter, but the deposit energy into grains is much lower.
According to \citet{herbst06}, the stochastic heating by protons leads to the maximum temperature of $\lesssim$20 K for 0.1 $\mu$m-sized dust with the frequency of once per $\sim$10 yr.
The temperature rise from 10 K to 20 K has a strong impact on the equilibrium \ce{H2} coverage \citep{furuya19}.
However, given much shorter timescale for the \ce{H2} coverage to reach the adsorption-desorption equilibrium ($\sim$1 ($10^4$ cm$^{-3}$/$\num{H_2}$) yr) than the interval of stochastic heating by protons, it is expected that for most of time, the \ce{H2} coverage is in the adsorption-desorption equilibrium at the local temperature and density.
Stochastic heating by irons is less relevant to the \ce{H2} coverage, because of much lower frequency of once per $\sim$10$^5$ yr \citep{herbst06}.
Then CR heating would not significantly affect the \ce{H2} coverage in the dense ($\gtrsim$10$^4$ cm$^{-3}$) ISM.

\renewcommand{\thefigure}{B\arabic{figure}}
\renewcommand{\thetable}{B\arabic{table}}
\setcounter{figure}{0}
\setcounter{table}{0}
\section{Detailed discussion for selected species} \label{sec:detail}
This appendix provides a more detailed explanation for species in Fig. \ref{fig:fullgrid_minor}. 
I describe the differences in the molecular abundances between the $\re$ model and the $\rembs$ model with $\sigma_i = 0.2 \overline{\edes_i}$ and [$\edes_{{\rm max},\,i}$, $\edes_{{\rm min},\,i}$] = [$\overline{\edes_i} + 3\sigma_i$, $\overline{\edes_i} - 3\sigma_i$], i.e., the model with the widest binding energy distribution in this work.

{\bf \ce{O2} ice}: the \ce{O2} ice abundance in the $\rembs$ model is orders of magnitude higher than that in the $\re$ model at $\leq$16 K.
In the $\rembs$ model, \ce{O2} ice is mainly formed by the recombination of two atomic O on the surface, while \ce{O2} is mainly formed in the gas phase and adsorbs onto dust grains in the $\re$ model. 
Both in the $\re$ and $\rembs$ models, the main destruction pathway of \ce{O2} ice is the hydrogenation reaction, which eventually leads to the \ce{H2O} formation \citep{miyauchi08,ioppolo08}.
As seen in Fig. \ref{fig:hop_coeff}, the relative hopping rate of atomic O with respect to H decreases with increasing temperature.
As a result, \ce{O2} ice abundance decreases with temperature.

{\bf \ce{H2S} ice}: the abundances of \ce{H2S} ice is higher in the $\rembs$ model compared to the $\re$ model at $\leq$16 K.
In the $\re$ model, at $\leq$16 K, the chemical desorption of \ce{H2S} occurs efficiently, 
through the \ce{H2S}–HS loop; HS is formed from \ce{H2S} via a hydrogen abstraction reaction, and hydrogenation of HS leads to the (re)formation of \ce{H2S}.
\ce{H2S} can desorb with a probability of $\sim$3 \% by chemical desorption in each cycle of this \ce{H2S}–HS loop \citep{oba18,furuya22a}.
In the $\rembs$ model, atomic H tends to be populated in deep potential sites, slowing down reactions relevant to atomic H on grain surfaces, leading to higher \ce{H2S} ice abundance.

{\bf \ce{SO} ice}: Both in the $\re$ and $\rembs$ models, \ce{SO} is mainly formed in the gas phase and frozen out onto dust grains. In the $\rembs$ model at 10 K, the reaction between atomic O and HS on dust grains also contributes to the SO ice formation.
SO ice is mainly destroyed by the reaction with atomic O, producing \ce{SO2} ice.
In the $\re$ model, at $\geq$16 K, thermal hopping of atomic O becomes efficient and the SO ice abundance decreases, increasing \ce{SO2} ice abundance.
In the $\rembs$ model, atomic O tends to be trapped in deep potential sites at $\geq$12 K, suppressing the destruction of SO ice (and the formation of \ce{SO2} ice).

{\bf \ce{SO2} ice}: in $\re$, \ce{SO2} ice is mainly formed by SO + O at $\ge$12 K, 
while at 10 K, \ce{SO2} is mainly formed in the gas phase and frozen out.
In $\rembs$, even at 20 K, \ce{SO2} is mainly formed in the gas phase, because of the reason described above. \ce{SO2} ice is mainly destroyed by UV photodissociation by CR-induced UV. As a result, the \ce{SO2} ice abundance is almost independent of the temperature.

{\bf \ce{OCS} ice}:  Both in the $\re$ and $\rembs$ models, OCS is formed by CO + S on grain surfaces, and destroyed by OCS + H $\rightarrow$ OC + HS on grain surfaces.
Because atomic S abundance on grains is higher in the $\rembs$ model than in the $\re$ model due to slower hydrogenation reactions, and because of the enhanced mobility of atomic S, the OCS ice abundance is higher in the $\rembs$ model than in the $\re$ model.

\renewcommand{\thefigure}{C\arabic{figure}}
\renewcommand{\thetable}{C\arabic{table}}
\setcounter{figure}{0}
\setcounter{table}{0}
\section{Effect of numerical resolution} \label{sec:comparison}
\begin{figure*}[ht!]
\plotone{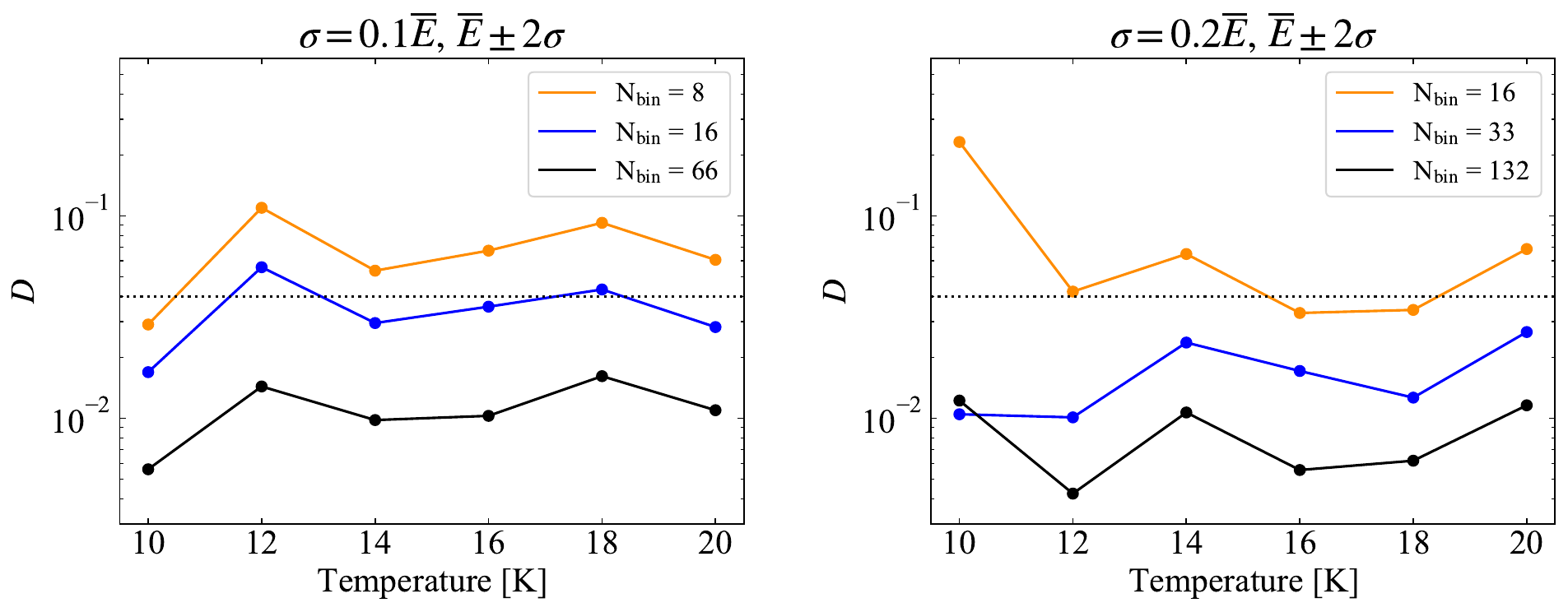}
\caption{
A pseudo-distance (Eq. \ref{eq:dist}) between the $\rembs$ model with $\nbin$ = 33 (left panel) or 66 (right panel) and the model with different $\nbin$. 
Binding energy distribution for each species is assumed to be a Gaussian distribution with [$\edes_{{\rm max},\,i}$, $\edes_{{\rm min},\,i}$] = [$\overline{\edes_i} + 2\sigma_i, \overline{\edes_i} - 2\sigma_i$].
$\sigma_i$ is set to $0.1\overline{\edes_i}$ and $0.2\overline{\edes_i}$ in the left and right panels, respectively. 
The horizontal dotted lines represent $D \sim 0.04$, which corresponds to the abundance difference of $\lesssim$10 \% on average.
The gas density is set to 10$^4$ cm$^{-3}$ and $A_V$ is set to 10 mag, varying the temperature from 10 K to 20 K. The abundances at 10$^6$ yr are used for the comparisons. 
}
\label{fig:diff_nbin}
\end{figure*}

To use the $\rembs$ method, one has to choose the number of bins for discretizing the binding energy distribution ($\nbin$).
Here the minimum value of $\nbin$ required to obtain converged results is discussed.
For this sake, the $\rembs$ models are additionally run with the same setting in Section \ref{sec:garrod}, varying in $\nbin$.
Figure \ref{fig:diff_nbin} shows a pseudo-distance $D$ (Eq. \ref{eq:dist}) between models with different $\nbin$ values.
Binding energy distribution for each species is assumed to be a Gaussian distribution with [$\edes_{{\rm max},\,i}$, $\edes_{{\rm min},\,i}$] = [$\overline{\edes_i} + 2\sigma_i, \overline{\edes_i} - 2\sigma_i$].
In the left and right panels, $\sigma_i$ is set to $0.1\overline{\edes_i}$ and $0.2\overline{\edes_i}$, respectively.
$y^0_i$ is the abundance of species $i$ in the case with $\nbin$ = 33 (left panel) or 66 (right panel), and the icy species with a total (surface + mantle) abundance of larger than 10$^{-12}$ at 10$^6$ yr is considered.
These $\nbin$ values were adopted in Section \ref{sec:garrod}.

When $\nbin$ is 33 or larger, the predicted abundances are similar, regardless of $\nbin$ ($D \lesssim 0.04$ which corresponds to the abundance difference of $\lesssim$10 \% on average).
In the case when $\sigma_i = 0.1\overline{\edes_i}$, the models with $\nbin = 16$ and $\nbin \geq 33$ give similar abundances. 
\citet{grassi20} noted that the result obtained with $N_{\rm bin} = 21$ is similar to that obtained with larger $N_{\rm bin}$.
Although $\nbin$ required to obtain converged results seems to depend on assumed binding energy distributions, physical conditions, and chemical network, at least $N_{\rm bin}$ of $\sim$20 seems to be necessary.


\renewcommand{\thefigure}{D\arabic{figure}}
\renewcommand{\thetable}{D\arabic{table}}
\setcounter{figure}{0}
\setcounter{table}{0}
\section{Additional figures} \label{sec:diff_model}
\begin{figure*}[ht!]
\plotone{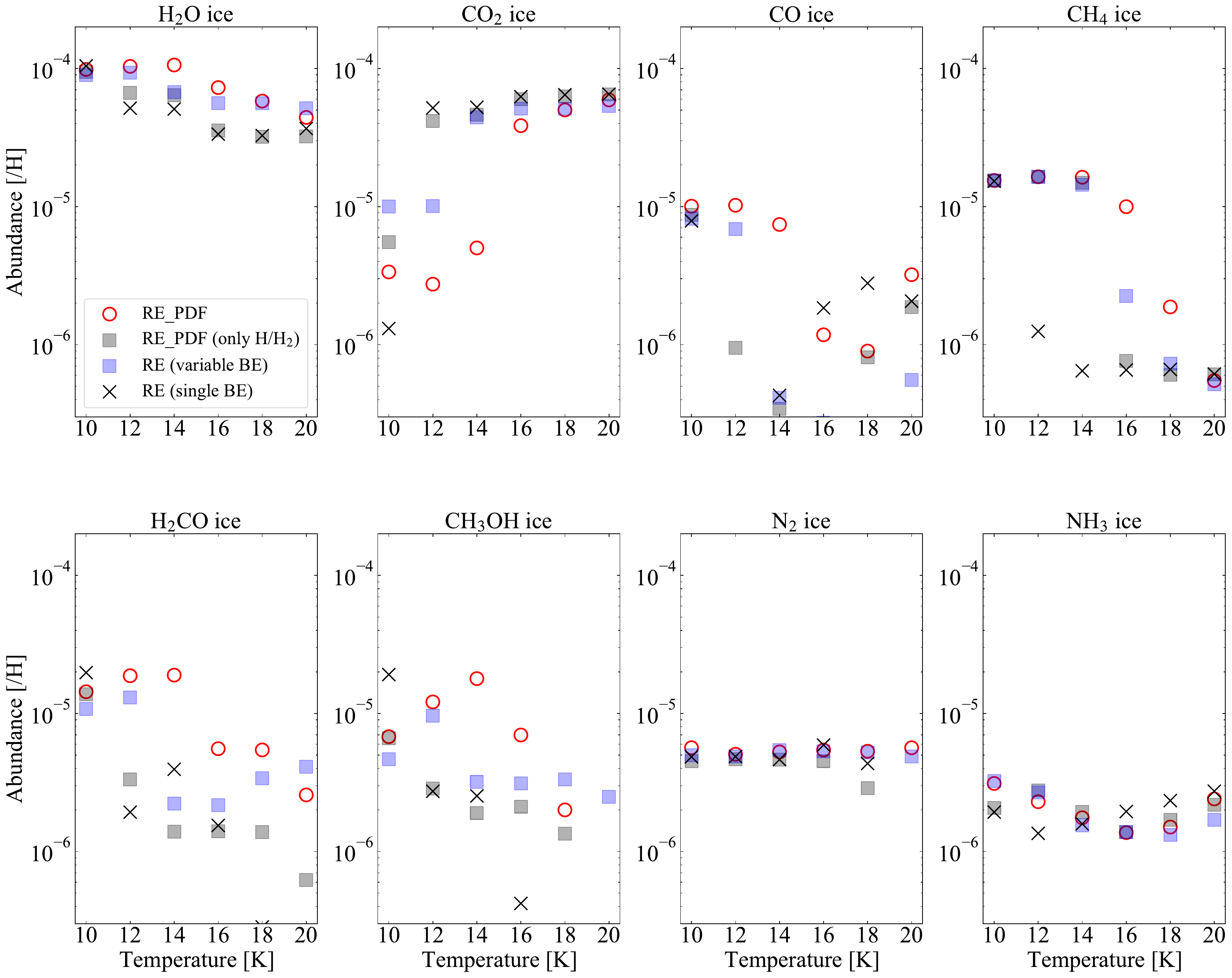}
\caption{
Abundances of selected icy species at 10$^6$ yr predicted by $\rembs$ (red circles), $\rembs$ only for atomic H and \ce{H2}, $\re$ with variable binding energy, and conventional $\re$ (crosses) at the density of 10$^4$ cm$^{-3}$ and $A_V$ = 10 mag, varying the temperature from 10 K to 20 K. Binding energy distribution for each species is assumed to be a Gaussian distribution with [$\edes_{{\rm max},\,i}$, $\edes_{{\rm min},\,i}$] = [$\overline{\edes_i} + 2\sigma_i, \overline{\edes_i} - 2\sigma_i$] and $\sigma_i = 0.2\overline{\edes_i}$.
}
\label{fig:grid_diff}
\end{figure*}

\begin{figure*}[ht!]
\plotone{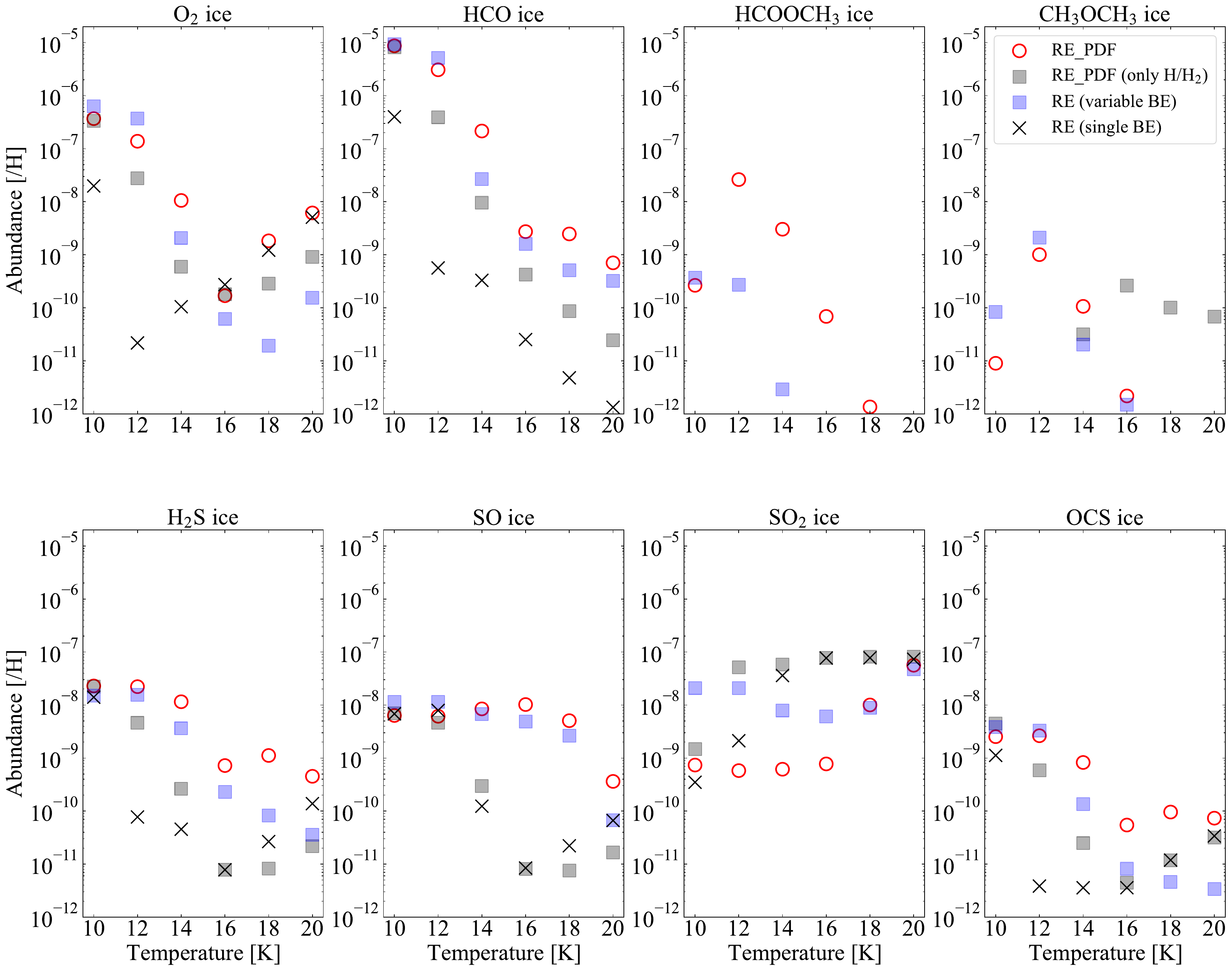}
\caption{
Similar to Fig. \ref{fig:grid_diff} but for less abundant species.
}
\label{fig:grid_diff_minor}
\end{figure*}

Figures \ref{fig:grid_diff} and \ref{fig:grid_diff_minor} compares the abundances of selected icy species predicted by $\rembs$, $\rehmbs$, $\re\_{\mu}$, and $\re$. See Section \ref{sec:simplified} for details.

\end{appendix}

\bibliography{ms}{}
\bibliographystyle{aasjournal}



\end{document}